\documentclass[%
 reprint,
superscriptaddress,
 amsmath,amssymb,
 aps,
]{revtex4-2}

\usepackage{graphicx}
\usepackage{dcolumn}
\usepackage{bm}

\begin{document}

\preprint{APS/123-QED}

\title{Machine learning in nuclear physics at low and intermediate energies}

\author{Wanbing He}
\email{hewanbing@fudan.edu.cn}
\affiliation{Key Laboratory of Nuclear Physics and Ion-beam Application (MOE), Institute of Modern Physics, Fudan University, Shanghai 200433, China}
\affiliation{Shanghai Research Center for Theoretical Nuclear Physics, NSFC and Fudan University, Shanghai 200438, China}

\author{Qingfeng Li}
\email{liqf@zjhu.edu.cn}
\affiliation{School of Science, Huzhou University, Huzhou 313000, China}
\affiliation{Institute of Modern Physics, Chinese Academy of Sciences, Lanzhou 730000, China}

\author{Yugang Ma}
\email{mayugang@fudan.edu.cn}
\affiliation{Key Laboratory of Nuclear Physics and Ion-beam Application (MOE), Institute of Modern Physics, Fudan University, Shanghai 200433, China}
\affiliation{Shanghai Research Center for Theoretical Nuclear Physics, NSFC and Fudan University, Shanghai 200438, China}

\author{Zhongming Niu}
\email{zmniu@ahu.edu.cn}
\affiliation{School of Physics and Optoelectronic Engineering, Anhui University, Hefei 230601, China}

\author{Junchen Pei}
\email{peij@pku.edu.cn}
\affiliation{State Key Laboratory of Nuclear Physics and Technology, School of Physics,
Peking University, Beijing 100871, China}
\affiliation{Southern Center for Nuclear-Science Theory (SCNT), Institute of Modern Physics, Chinese Academy of Sciences, Huizhou 516000,  China}

\author{Yingxun Zhang}
\email{zhyx@ciae.ac.cn}
\affiliation{Department of Nuclear Physics, China Institute of Atomic Energy, Beijing 102413,  China}
\affiliation{Guangxi Key Laboratory of Nuclear Physics and Technology, Guangxi Normal University, Guilin 541004, China}

\date{\today}

\begin{abstract}
Machine learning is becoming a new paradigm for scientific research in various research fields due to its exciting and powerful capability of modeling tools used for big-data processing task. In this mini-review, we first briefly introduce different methodologies of the machine learning algorithms and techniques. As a snapshot of many applications by machine learning, some selected applications are presented especially for low and intermediate energy nuclear physics, which include topics on theoretical applications in nuclear structure, nuclear reactions, properties of nuclear matter as well as experimental applications in event identification/reconstruction, complex system control and firmware performance. Finally, we also 
give a brief summary and outlook on the possible directions of using machine learning in low-intermediate energy nuclear physics and possible improvements in machine learning algorithms.
\end{abstract}

\maketitle


\section{Introduction}\label{section1}
%

In the past decade, Machine Learning (ML) or Artificial Intelligence (AI) driven applications have been extensively and successfully exhibited in many different fields~\cite{LeCNt521,JorSc349,MEHTA2019PR}. Especially, the advents of AlphaGo~\cite{alphago} and AlphaFold~\cite{alphafold} have deeply changed the society's view about ML and AI.  ML is a branch of AI and focuses on the use of data. The application of ML has a long story. This revival wave of ML is different from previous ones, due to fast developments of ML algorithms and computing infrastructures, high-speed data transmission, and strong driving forces from commercial and industrial applications  in the digital era. In particular, the deep learning (DL) techniques have demonstrated powerful capabilities. The number of ML papers published in physics journals has grown rapidly since the year 2016. There is no reason that scientists should ignore the opportunities brought by ML~\cite{RevModPhys2022}. 

Modern scientific research is continuously producing massive data. The treatment of complex, heterogeneous, imperfect and
high-dimensional correlated data is beyond conventional numerical methods and statistical methods. 
ML is promising in this respect by providing new data-driven methodologies. 
The idea of physics model is by resolving the problems using fundamental laws and equations of physics, which are simplified, predictable, 
and involve few data. 
However, the realistic problems are usually very complex and involve plenty of inputs and outputs, although
correlations between them may be weak or indirect. 
The idea of data model is rooted in the belief that data includes all the correlations and rules.
ML can learn existing data information and infer if new data follows the same data correlations and rules. 


Compared to physics models, the advantage of ML is to solve unconventional complicated problems that are difficult to be described by equations.  The weakness of ML is the black-box modeling of data with
little physics information. To this end, the physics knowledge can be used as a prior in Bayesian learning.
The merger of ML and physics has attracted great interests and can facilitate  
 more effective learning and  more reliable inference. 
There are a number of attempts to develop physics-informed, or physics-guided, or physics-constrained ML~\cite{Carleo2019,MumpowerPRC2022,Karniadakis2021}.
Physics knowledge can also be used to transform data and provide additional input information. 

\begin{figure}[htbp]
\centering
\includegraphics[scale=0.45]{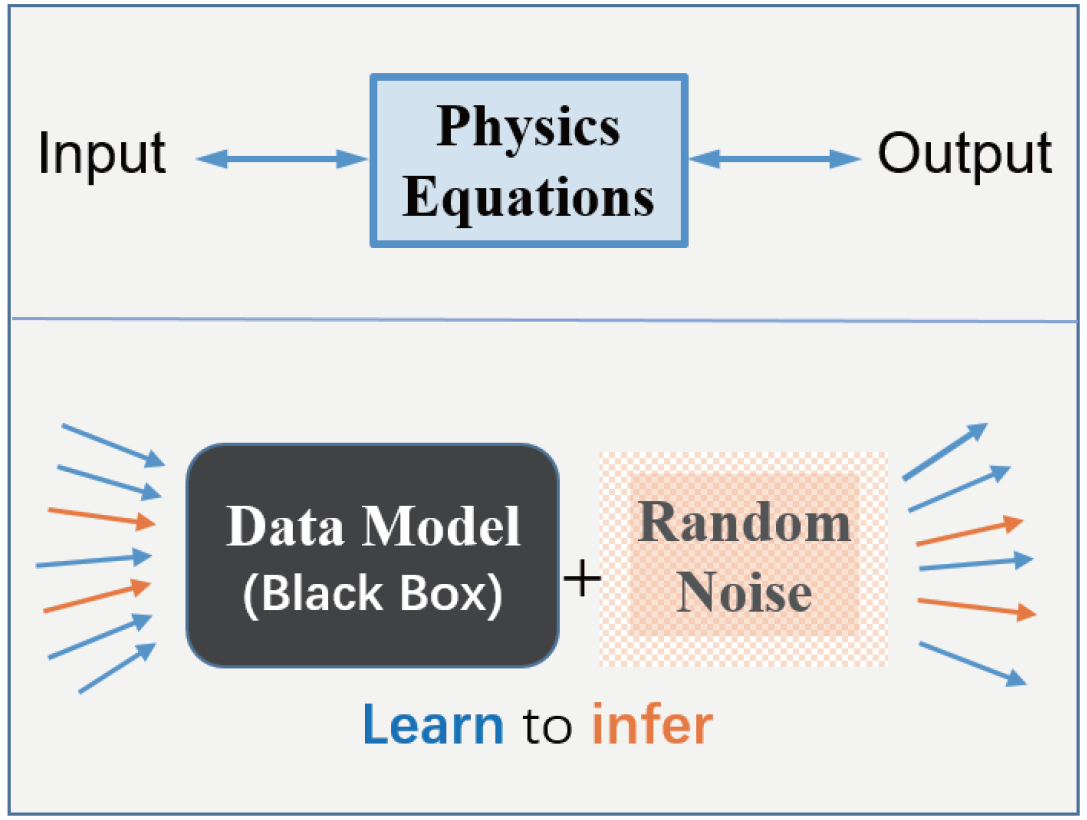}
\caption{(Color online) The schematic figure shows the difference between physics model and data model.
}
\label{fig:example2}
\end{figure}

The capabilities of ML in research are rather broad. ML is often used for classification and regression. For example, face recognition is a typical classification problem and is now a very successful commercial usage of ML. As a classification problem, ML can be used to improve the particle identification and event reconstruction in nuclear and particle physics experiments by exploiting a large feature space. The regression is to solve ill-inverse problems with uncertainties and infer from existing data. ML with underlying data correlations as an alternative approach can be used for predicting nuclear properties at unknown regions.

In addition to basic applications, ML has been used for emulations~\cite{keeble2020,ZhangXL2022},
model mixing~\cite{NazarewiczPRC2020}, data fusion~\cite{WangZA2022}, optimizations~\cite{EkstromJPG2019},  model reductions~\cite{ZhangR2019PRB} and uncertainty quantifications~\cite{Band2021}. 
For expensive computing problems and experiments, the emulations by ML
can significantly alleviate the costs. Once the complex problems are emulated, or better interpolated, the control, design, and optimizations can be facilitated. The emulation is different from simulation, such as Monte Carlo simulation, which 
is forward and based on physics rules. The emulation is also different from fitting, which usually
refers to problems with a few variables and with assumed functions. 
Neural networks can approximate  many body wave functions to speed up calculations.
Different models are based on different perspectives, then a comprehensive inference by model mixing
should be more reliable~\cite{NazarewiczPRC2020}. 
To evaluate imperfect data, the data-fusion can provide more accurate and useful information by
including underlying correlations than separated data, so that
the maximum values of noisy, discrepant and incomplete nuclear data are exploited~\cite{WangZA2022}.

ML is not restricted to a novel data tool, it is also useful in theoretical and experimental studies~\cite{Karagiorgi2022}.
It can help to infer nuclear equation of state from multi-message astrophysical observations and
terrestrial nuclear experiments~\cite{AnnalaPRX}. 
Neural networks can approximate the solutions to partial differential equations much faster than traditional numerical methods~\cite{Karniadakis2021}. It is expected to assist us to find new physics theories from scratch~\cite{Max2022PRL}.
In studies with insufficient data, synthetic data generated by models or rules can even be used to train for faster and accurate predictions.

The development of ML will undoubtedly bring great opportunities in nuclear physics studies.
The experiments on large rare-isotope beam facilities are important for our understandings of exotic nuclear properties far from
stability and  nuclear matter under extreme conditions. 
These experiments are expensive and very difficult, and they should be delicately planed, analyzed and explained, in which ML can play an essential role. 
Furthermore, there are some nuclear regions that are of interests for nuclear astrophysical processes, are almost impossible to access by experiments.
On the other hand, it is not feasible for $ab~initio$ calculations to reach heavy nuclei or dense nuclear matter. 
The combination of  multi-messenger  observations and various theories by machine learning is expected to provide better inferences and evaluations~\cite{AnnalaPRX}. 
It is also possible to reduce the costs of $ab~initio$ computations by machine learning in the future~\cite{Adams2021PRL}.

In this review, the applications of ML in low and intermediate energy nuclear physics in recent years are reviewed, to boost further developments in this field. 
Several typical methodologies in ML are explained in the following section~\ref{ML-methodology}. 
Then some selected applications of ML in nuclear physics are discussed in details in section~\ref{Applications}.
Finally this review is concluded with a summary and an outlook. 

\section{Methodology of Machine learning}
\label{ML-methodology}

Generally, the ML algorithms are used to learn the mapping relationship between input variables and output variables from data automatically. Thus, the common architecture of ML will consist of three ingredients. The first thing we need is the data set for training and validation, the second is to build a relation between inputs and outputs, and the third is the algorithm to optimize the hyperparameters of ML algorithms $\theta$ by estimating the performance which are usually selected as mean square-error.

More details, a set of $n$ samples of data with $X_\mu$ and label $y_\mu$ is provided, i.e., the data set $D = \{(X_\mu,y_\mu)_{\mu=1,\cdots, n}\}$. In physics studies, $X_\mu$ could be a series of input variables of the theoretical model, $y_\mu$ could be an experimental data corresponding to the input variables. The goal of supervised learning is to find a function $f$ to map $X_\mu$ to $y_\mu$, i.e., $y_\mu = f(X_\mu,\theta)$, based on the provided data. In practical calculations, the dataset $D$ is usually called the training set and usually splits the available data samples into the training set and validation set. To learn the function $f$, one can optimize $\theta$ by evaluating the performance. We define $\theta^*$ as the best parameter in ML algorithms after optimization. Once $f$ is constructed based on the data set, a new sample $X_{new}$ without its label can be approximated well with the label by the output of the function $f(X_{new},\theta^*)$.

The frequently used ML algorithms is neural network. According to its architecture, it can be divided into Artificial Neural Network (NN)~\cite{Berndt2012}, Bayesian Neural Network (BNN)~\cite{Neal1996}, Convolutional Neural Network (CNN), Kernel Ridge Regression (KRR), and so on, for the task of regression and classification.



\textit{Neural Network}: In a simple neural network, the function $f$ is expressed in terms of a set of network parameters $\theta = \{a,b_j,c_j, d_{ij}\}$, i.e., $f(X_\mu, \theta)$,
\begin{equation}
	f(X_\mu, \theta)=g(\theta\cdot X_\mu)=a+\sum_{j=1}^H b_j g_j\left(c_j+\sum_{i=1}^l d_{ij}x_i\right)
\end{equation}
$H$ and $l$ denotes neurons in each layer, and the model parameters (or “connection weights”)
are $\theta = \{a,b_j,c_j, d_{ij}\}$. $g_j$ is an active function, which can be chosen as $g_j = tanh$, $sigmoid$, $ReLU$, and so on. A loss function $\mathcal L$ is used to quantify the distance between $f(X_\mu, \theta)$ and $y_\mu$. For all data, the average of the loss over the training set is needed, which is called the empirical risk $R(f(X_\mu,\theta))=\frac{1}{n}\sum_{\mu=1}^{n} \mathcal L$. The training procedure is to adjust the network parameter $\theta$ to minimize the empirical risk.

To minimize the empirical risk function over the weights, a gradient descent algorithms was most commonly used. In this method, $\theta$ is iterated as, 
\begin{equation}
	\theta^{t+1} = \theta^t-\gamma \nabla_\theta \mathcal R(f_\theta),
\end{equation}
where $\gamma$ is the learning rate, and the weights $\theta$ are iteratively adjusted in the direction of the gradient of the empirical risk. There are many variants of the gradient descent, such as the stochastic gradient descent (\textit{SGD}) where the full empirical risk function $R$ is replaced by the contribution of just a few of the samples, \textit{Adam} where the learning rate is adaptive momentum estimation, \textit{Adagrad} which is a family of sub-gradient algorithms for stochastic optimization. Consequently, these approaches to neural networks are  to find a single best-fit value $\theta^*$ for the neural network parameters, and hence a single best-fit neural network, $f (X_\mu,\theta^*)$.

The training error measures how well such a minimization is achieved, and they depend on the size of data sample, optimization algorithm, and the approximation \cite{EWN}. Suppose that the $\hat f$ is the output from the machine learning algorithm, $f_m$ is the best approximation in data space $\mathbb R$, $\tilde{f}_{n,m}$ is the best approximation of data set sample $\mathbb S$, thus, the errors can be divided into three parts, 
\begin{equation}
	f-\hat f = \underbrace{f-f_m}_{appr.}+\underbrace{f_m-\tilde f_{n,m}}_{estim.}+\underbrace{\tilde f_{n,m}-\hat f}_{optim.}.
\end{equation}
The first term depends on the algorithm for finding the minimization of empirical risk, the second and third error depend on both the size of sample and optimization algorithm. Thus, improving the accuracy of learning should be realized by quality of data sample, by randomly picking a fraction of the available data which are tested via using different optimization algorithm.

For deep learning, the layers of network are much more than the simple neural network, which is essentially a neural network with three or more layers. For a $L$-layer fully connected neural networks, the function $f(X_\mu,\theta)$ is parameterized as follows~\cite{CarleoRMP2019}:
\begin{equation}
	f(X_\mu) = g^{(L)}\left(\theta^{(L)},\cdots g^{(2)}\left(\theta^{(2)},g^{(1)}(\theta^{(1)},X_\mu)\right)\right).
\end{equation}

Another important and powerful variant of deep neural networks are convolutional neural networks (CNN). In the CNN, the
input into each of the hidden units is not whole, but is obtained via a convolutional kernel which is a filter applied to a small part of the input space. The filter will be shifted to different positions corresponding to different hidden units. Compared to the fully connected neural networks, each layer of the CNN has a much smaller number of parameters, which is in practice advantageous for the learning algorithms. CNN is usually used for analysis of images. 

For neural network, one can also use it to solve the partial differential equation (PDE). 
To obtain the approximative solution of a PDE via deep learning, a key step is to construct a neural network $f(X_\mu,\theta)$ to minimize the PDE residual, $\mathcal{R}$. The neural network is named as physics-informed neural networks (PINN)~\cite{LuLu2021,Karniadakis2021}. In PINNs, the derivatives of the network outputs $f$ with respect to the network inputs $X_\mu$ is calculated by automatic differentiation (AD, also called algorithmic differentiation), which are evaluated using back-propagation, a specialized technique of AD. Compared to the traditional mesh-based methods, such as the finite difference method (FDM) and the finite element method
(FEM), deep learning could be a mesh-free approach by taking advantage of the automatic differentiation, and could break the curse of dimensionality. 


The Bayesian Neural Network (BNN) is another important variant of neural network, which is constructed based on Bayes’ theorem. In BNN, a connection between a probability of given set of data $D$ and a given hypothesis (or model) or model parameter $\theta$ is provided as follows, 
\begin{equation}\label{bayes}
	p(\theta|D) = \frac{p(D|\theta)p(\theta)}{p(D)}.
\end{equation}
The posterior probability $p(\theta|D)$ is the probability of the assumed hypothesis or model parameters at given data $D$, and $p(\theta)$ is the prior probability of the hypothesis or model parameters $\theta$, $p(D|\theta)$ is the likelihood function. In practice of BNN, the hypothesis or model parameters refer to the network parameter $\theta$ and $D$ will be composed of the interested input data and output data, i.e., $D\equiv(X_\mu, y_\mu)$. The standard practice for likelihood function is to assume a form of  Gaussian distribution which is based on an objective (or ``loss'') function obtained from a least-squares fit to the empirical data. That is,
\begin{equation}
	p(D|\theta) = \exp(-\chi^2(\theta)/2), \chi^2=\sum_{i}^N \frac{[y_i-f(X_i,\theta)]^2}{\Delta y_i^2}, 
\end{equation}
where $N$ is the number of empirical data, $y_i \equiv y (X_i)$ is the i$th$ observable with its associated error $\Delta y_i$, and the inference of $y_i$ from the function $f(X_i,\theta)$ depends on both the input data $X$ and the network parameters $\theta$. The posterior distributions are obtained by learning the given data. With new data $X_n$, the average values of $f_n$, i.e., $\langle f_n \rangle$ will be obtained by integrating the neural network over the posterior probability density of parameters $p(\theta|X,y)$,
\begin{equation}\label{avfn}
	\langle f_n\rangle = \int f(X_n,\theta)p(\theta|X,y)d\theta.
\end{equation}
The high-dimensional integral in Eq. (\ref{avfn}) is approximated by Monte Carlo integration in which the posterior probability $p(\theta |X, y )$ is sampled using the Markov chain Monte Carlo method.

The prior probability $p(\theta)$ should be obtained with what is known about the neural network parameters. Usually, the form of prior is taken as a zero mean Gaussian prior for each neural network parameter, and the precision is presented as inverse of variances of these Gaussian distributions. By setting them as gamma distributions, the precision can vary over a large range and hence the BNN approach can search the optimal values of precision in the sampling process automatically.

Another classical supervised learning methods are based on so-called decision trees, support vector machine, Bayesian inference and naive Bayesian probability classifier. We briefly introduce them in the following part.

\textit{Decision Tree}: The decision tree has a tree structure, which consists of a root node, branches, internal nodes and leaf nodes, and predicts the label $y$ of leaf which is associated with an instance $x$, i.e., $h : X \to Y$ by traveling from a root node of a tree to a leaf. At each node on the root-to-leaf path, the successor child is chosen on the basis of a splitting of the input space. Usually, the splitting is based on one of the features of $x$ or on a predefined set of splitting rules. It is used to go from observations about a data sample (represented in the branches) to conclusions about the item’s target value (represented in the leaves). There are also many variants of decision tree algorithms, i.e., LightGBM (Light Gradient Boosting Machine) and XGBoost (eXtreme Gradient Boosting).

\textit{Support vector machine}: 
The support vector machine (SVM) algorithm described in Refs.~\cite{Vapnik1995,Vapnik1998,William2007} is mainly used to to the problem of classification as well as for neural networks. SVMs belong to supervised learning and are generally easier to implement than neural nets. For a given training data set $D = \{(X_\mu,y_\mu)_{\mu=1,\cdots, n}\}$, in which $X_\mu$ is the data point and $y_\mu$ is a label. SVM will seek a hyper-plane to distinguish different labels of data. More details can  be found in Ref.~\cite{William2007}.

\textit{Bayesian inference and Naive Bayesian probability method}:
The so-call Bayesian inference is a statistical tools for analyzing the relation between model parameters and data. As in Eq.~(\ref{bayes}), it is also constructed based on Bayes’ theorem. But the differences are the selection of $D$ and $\theta$. In the Bayesian inference, $D$ usually is selected as the data or observations, i.e., $D = {y_i}$. $\theta$ is the physical model parameters rather than network parameters, i.e., $\theta = {x_i}$. 
The posterior probability $p(x|y)$ is the probability that the model parameters is true given data $y$, and $p(x)$ is the prior probability of the model parameters $x$. The likelihood function $p(y|x)$ has a form of  Gaussian distribution as in BNN, but the $\chi^2$ only depends on model parameter $x$. 

The naive Bayesian probability (NBP) classifier is also rooted in Bayesian theorem, but $\theta$ is the value of class variable and $D$ is dependent feature. Thus, Bayes theorem states the relationship between given class variable $\theta$ and dependent feature vector $D$. The so-called naive assumption is conditional independence between every pair of features.

In Table~\ref{tab:mlmethods}, we listed the typical machine learning algorithm used in this paper. The fourth column is the type of learning, which are usually divided into three broad categories: supervised learning, unsupervised learning, and reinforcement learning. Supervised learning concerns learning from labeled data, and the corresponding tasks include classification and regression. Unsupervised learning is concerned with finding patterns and structure in unlabeled data, they are usually used as clustering, dimensionality reduction, and generative modeling. The reinforcement learning concerns learning by interacting with an environment and changing its behavior to maximize its reward. It has been used to control the facility, such as Magnetic control of tokamak plasmas \cite{Degrave2022Nature}, but few to nuclear physics and we will not discuss it in this review paper. One should keep in mind that the distinction between the three types of ML is sometimes fuzzy and fluid.

\begin{table*}[htbp]
	\caption {(Color online) Typical ML methods discussed in this review paper.}
	\begin{tabular}{p{2cm}cp{5cm}cp{2cm}}
		
		\hline
		\hline
		Method & Architecture & Explanation & Learning type & Applications\\		
		\hline
		ANN & \begin{minipage}[b]{0.4\columnwidth}
			\centering
			\raisebox{-.5\height}{\includegraphics[width=\linewidth]{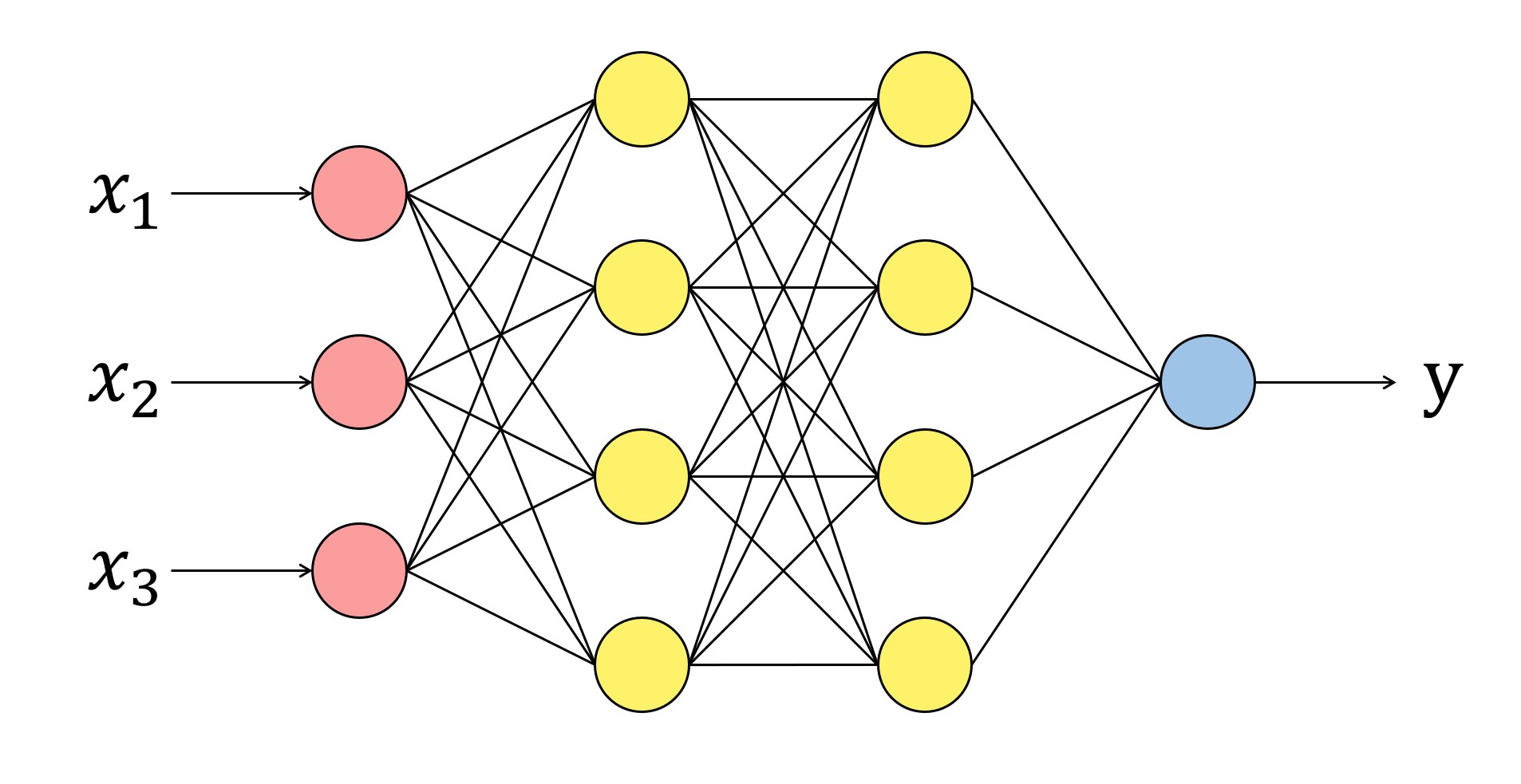}}
		\end{minipage} & Learning the network parameters $\theta$ to best fit $y_\mu=f(X_\mu,\theta)$ with training data & Supervised & Approximating function\\
		
		DNN & \begin{minipage}[b]{0.4\columnwidth}
			\centering
			\raisebox{-.5\height}{\includegraphics[width=\linewidth]{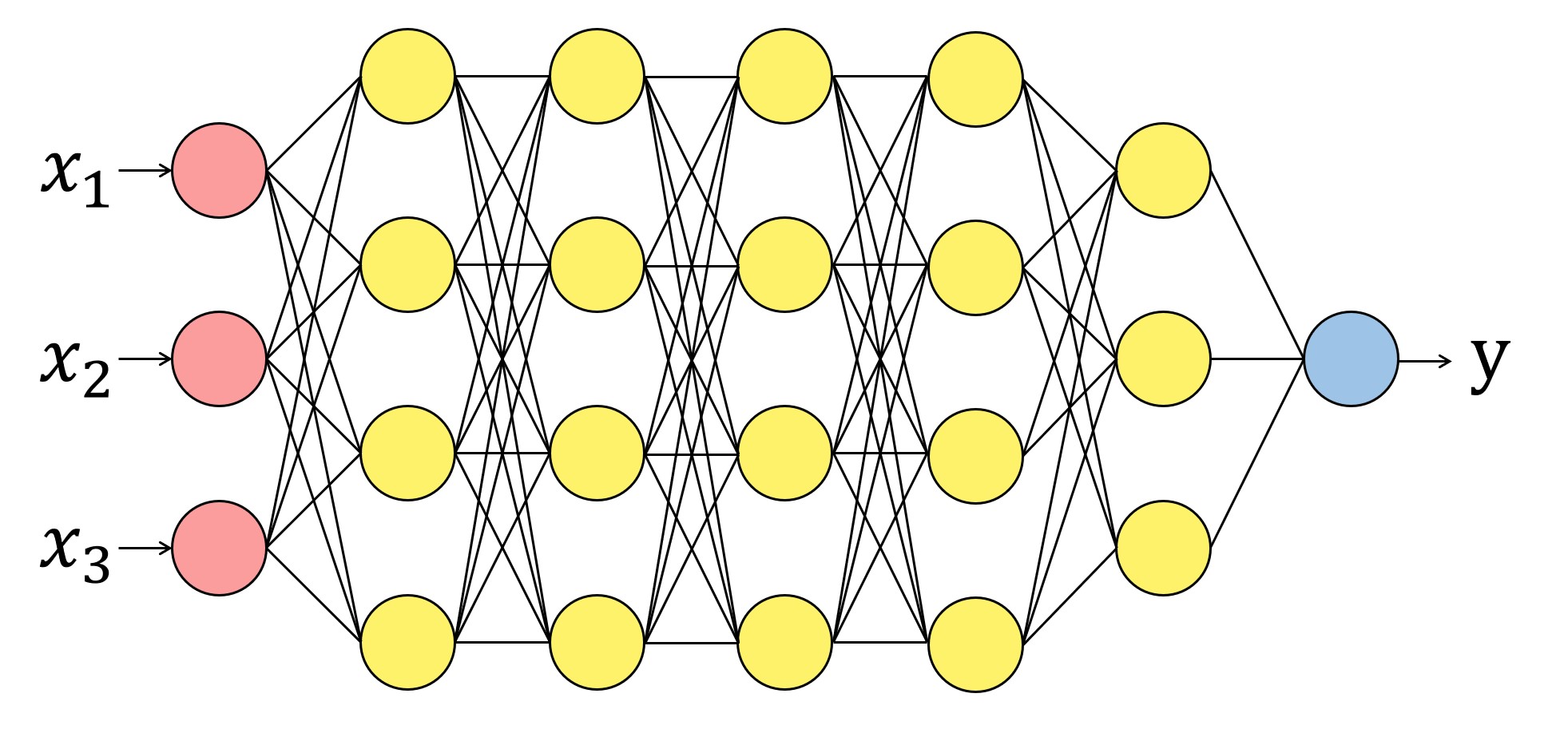}}
		\end{minipage} & Similar with ANN but with more layers & Supervised & Approximating function  \\	
		
		BNN & \begin{minipage}[b]{0.4\columnwidth}
			\centering
			\raisebox{-.5\height}{\includegraphics[width=\linewidth]{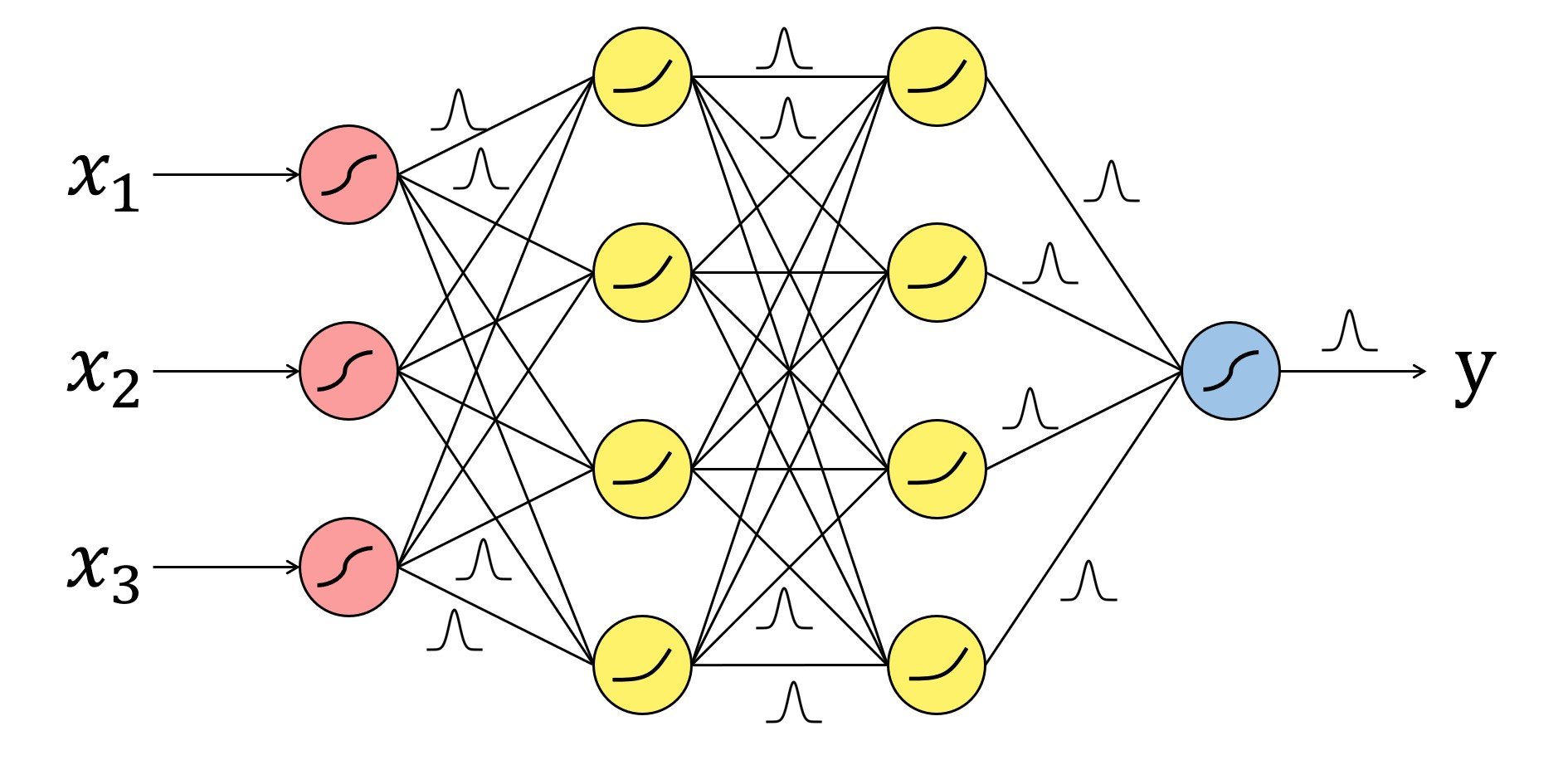}}
		\end{minipage}  & Similar with ANN but the net parameters have gaussian distributions, $P(\theta)\propto \mathcal{N}(\mu_{\theta}, \sigma_\theta)$ & Supervised & Approximating function \\
	
		CNN & \begin{minipage}[b]{0.4\columnwidth}
			\centering
			\raisebox{-.5\height}{\includegraphics[width=\linewidth]{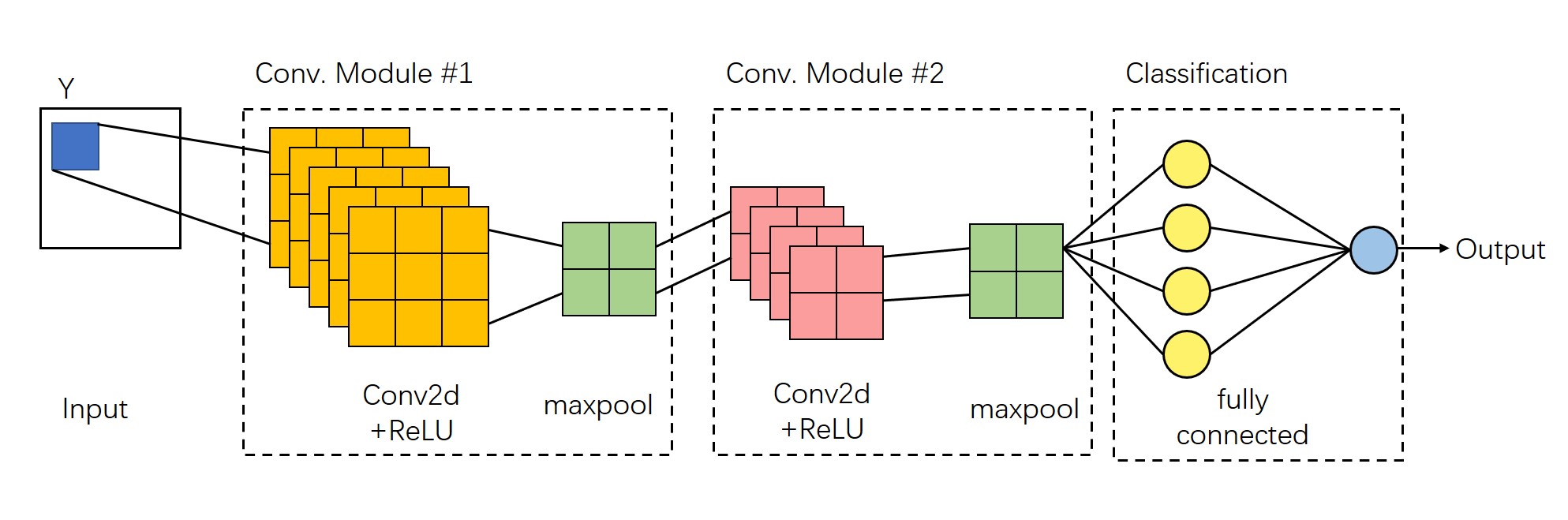}}
		\end{minipage} & A deep learning neural network designed for processing structured arrays of data & Supervised & Image classification \\

		GNN & \begin{minipage}[b]{0.4\columnwidth}
			\centering
			\raisebox{-.5\height}{\includegraphics[width=\linewidth]{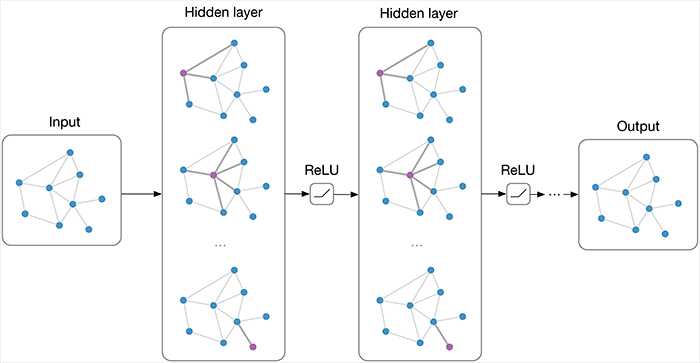}}
		\end{minipage} & Model the relationship between the nodes in a graph and produce a numeric representation of it & Supervised & Pattern classification \\

		RBF & \begin{minipage}[b]{0.4\columnwidth}
	\centering
	\raisebox{-.5\height}{\includegraphics[width=\linewidth]{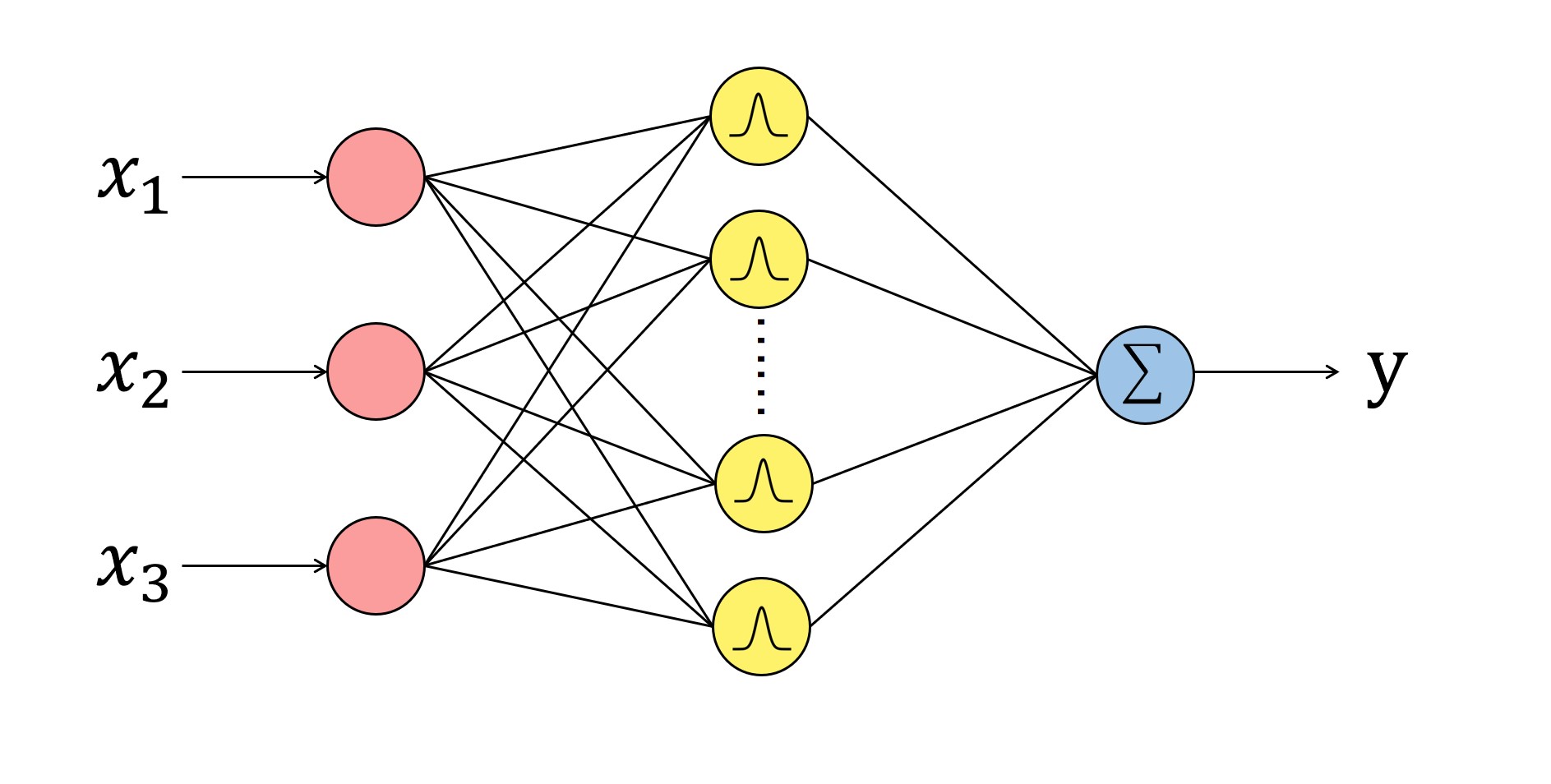}}
\end{minipage} & Three layers feed-forward neural networks with a Gaussian function as radial basis functions & Supervised & Approximating function \\

			
		VAE & \begin{minipage}[b]{0.4\columnwidth}
	\centering
	\raisebox{-.5\height}{\includegraphics[width=\linewidth]{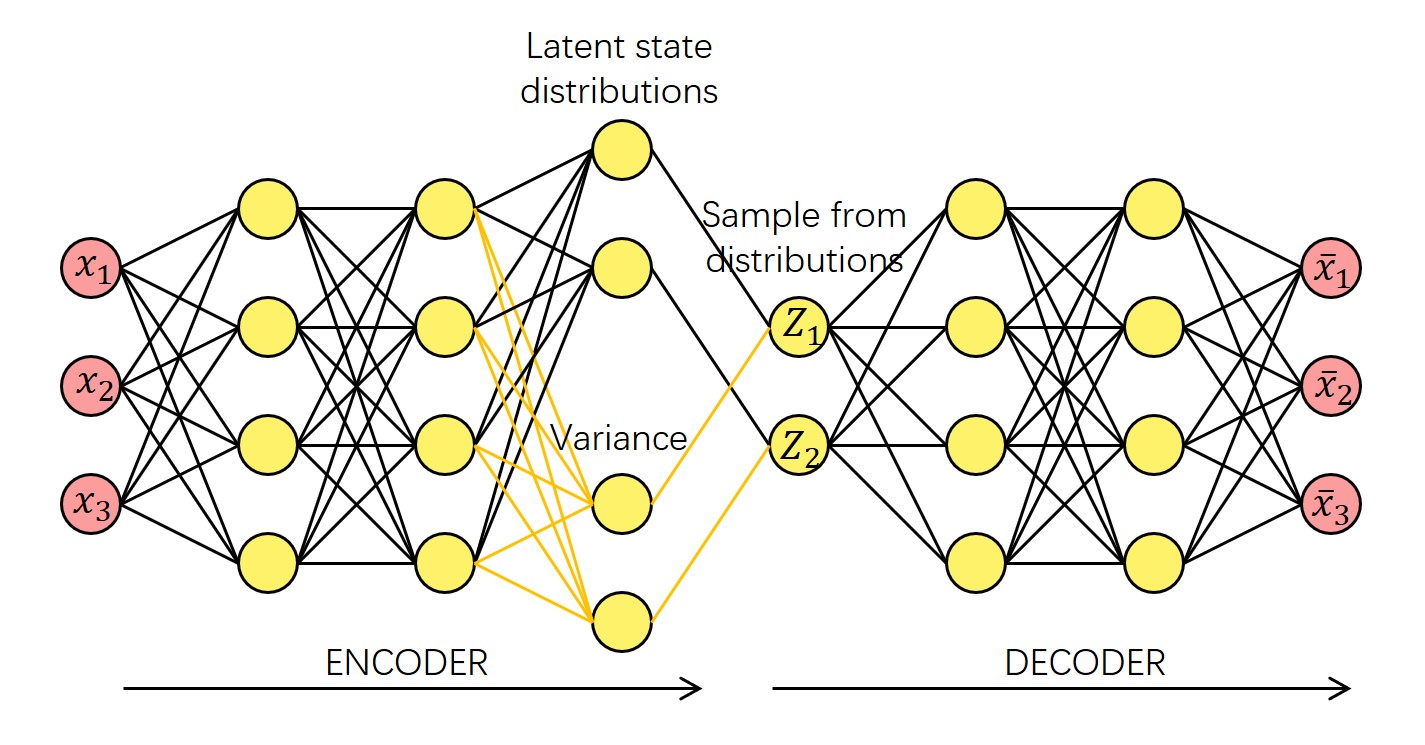}}
\end{minipage} & Provides a statistic manner for describing the samples of the dataset in latent space & Unsupervised & Describing an observation in latent space \\		

		SVM & \begin{minipage}[b]{0.4\columnwidth}
	\centering
	\raisebox{-.5\height}{\includegraphics[width=\linewidth]{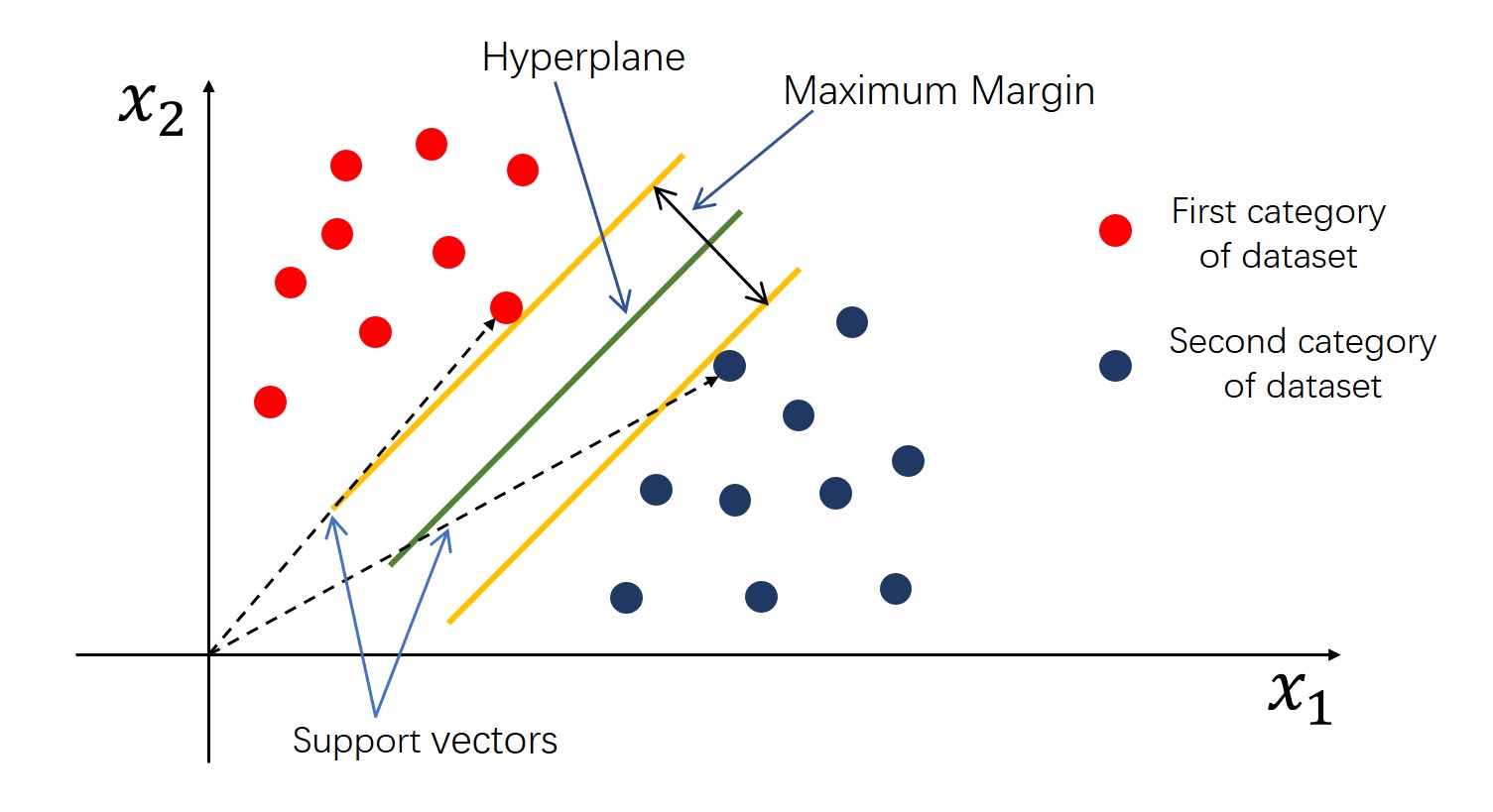}}
\end{minipage} & Finding a hyper-plane to distinguish different types of data & Supervised & Classification or Regression  \\

		LightGBM & \begin{minipage}[b]{0.4\columnwidth}
	\centering
	\raisebox{-.5\height}{\includegraphics[width=\linewidth]{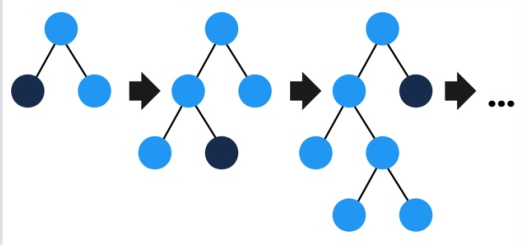}}
\end{minipage} & A gradient boosting framework that uses tree based learning algorithms & Supervised & Classification or Regression\\		

K-means & \begin{minipage}[b]{0.4\columnwidth}
	\centering
	\raisebox{-.5\height}{\includegraphics[width=\linewidth]{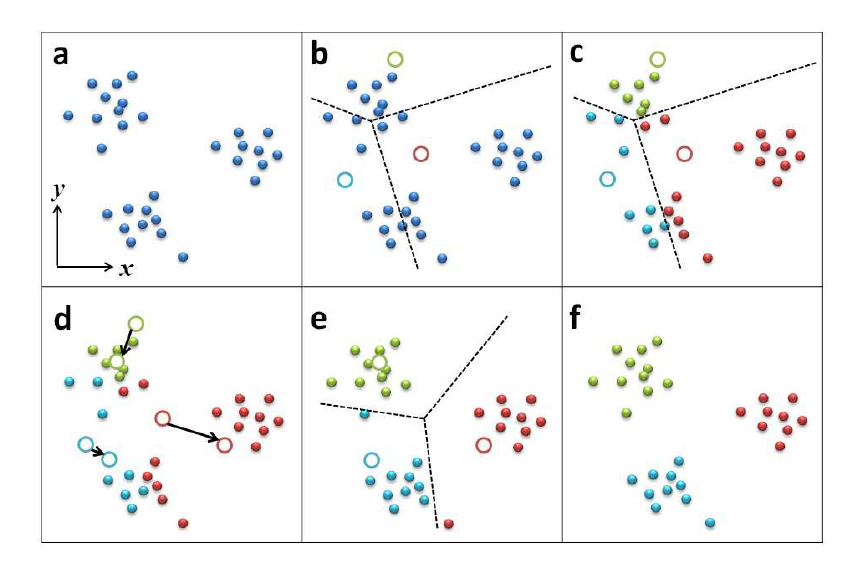}}
\end{minipage} & Grouping the unlabeled dataset into different clusters & Unsupervised & Detecting abnormal data, clustering \\		

Bayesian inference & \begin{minipage}[b]{0.4\columnwidth}
	\centering
	\raisebox{-.5\height}{\includegraphics[width=\linewidth]{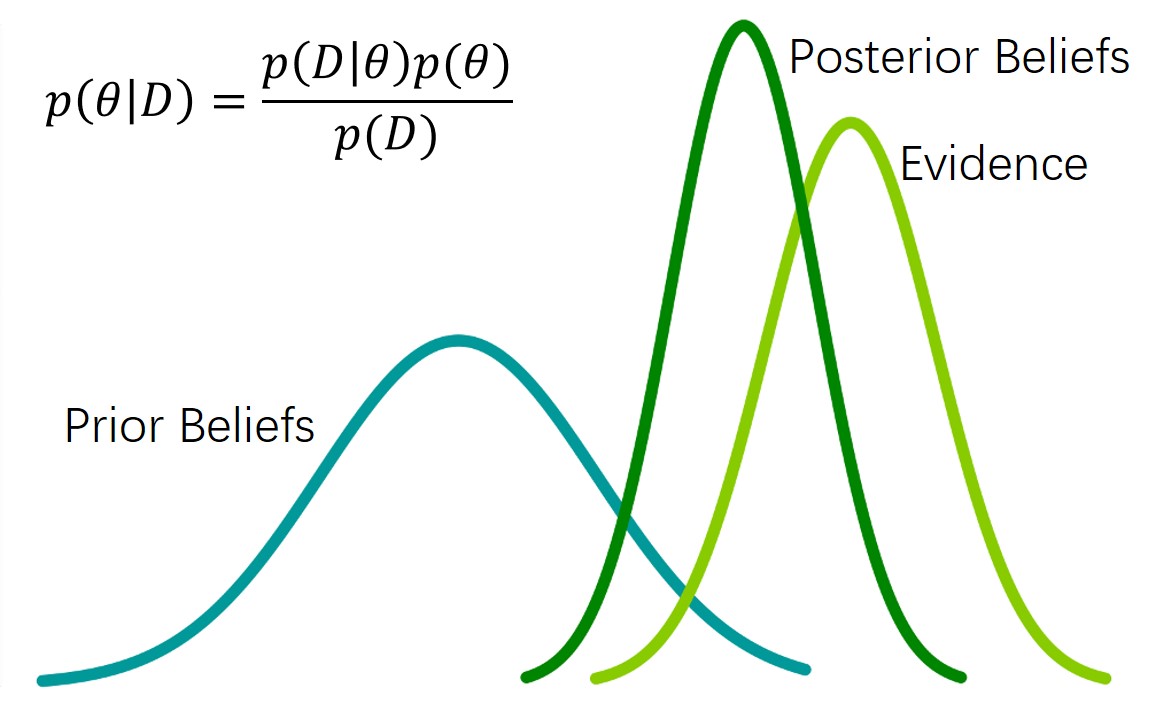}}
\end{minipage} & Inferences probabilities of model parameters with the prior information and related experiments & Supervised & Parameter selection and estimation \\		

		\hline
		\hline
		
	\end{tabular}
	\label{tab:mlmethods}
\end{table*}

In the end, we want to emphasize  that only part of ML algorithms is mentioned due to the scope of this review paper. More ML algorithms are suggested to find in the ML related textbooks. 





\section{Applications of machine learning in nuclear physics at low and intermediate energies}
\label{Applications}


\subsection{Nuclear structure} 
Nuclear structure is a basic and important research field in nuclear physics, which also plays an important role in both astrophysics and particle physics. The measurement of nuclear properties has made great progress in the past decades with the development of Radioactive Ion Beam (RIB) facilities. More and more nuclear structure data have been accumulated and evaluated, such as nuclear masses, charge radii, decay properties, and fission yields. However, the accurate prediction of nuclear properties remains a challenging and longstanding theoretical task due to the difficulties in the quantum many-body problem and the complexity of nuclear force. The machine learning provides a powerful and novel tool to learn and make predictions from data, whose applications in nuclear structure have grown rapidly during the past years~\cite{Bedaque2021EPJA, Boehnlein2022RMP}.

\subsubsection{Nuclear structure observables}
Nuclear mass is a fundamental quantity in nuclear physics. It contains a wealth of nuclear structure information such as magic number and shape transition, and has been widely used to extract nuclear effective interactions. It can be used to determine the energy release of a given reaction or decay process, so it is crucial to understand the origin of elements in the universe and energy generation in stars. With the development of radioactive ion beam facilities, more than 2000 nuclear masses have been measured whose accuracies are better than $100$ keV. The rms deviations between experimental data and theoretical nuclear mass predictions are generally around 500 keV, while the deviations among various mass models even reach tens of MeV when extrapolated to the unknown very neutron-rich region. Therefore, the current theoretical accuracies are still far from those required by the studies on exotic nuclear structure and astrophysical nucleosynthesis. The machine learning has been used to improve the accuracy of nuclear mass predictions, such as ANN, BNN, LightGBM, RBF, KRR, and NBP methods.

\begin{figure}[htbp]
\centering
\includegraphics[scale=0.26]{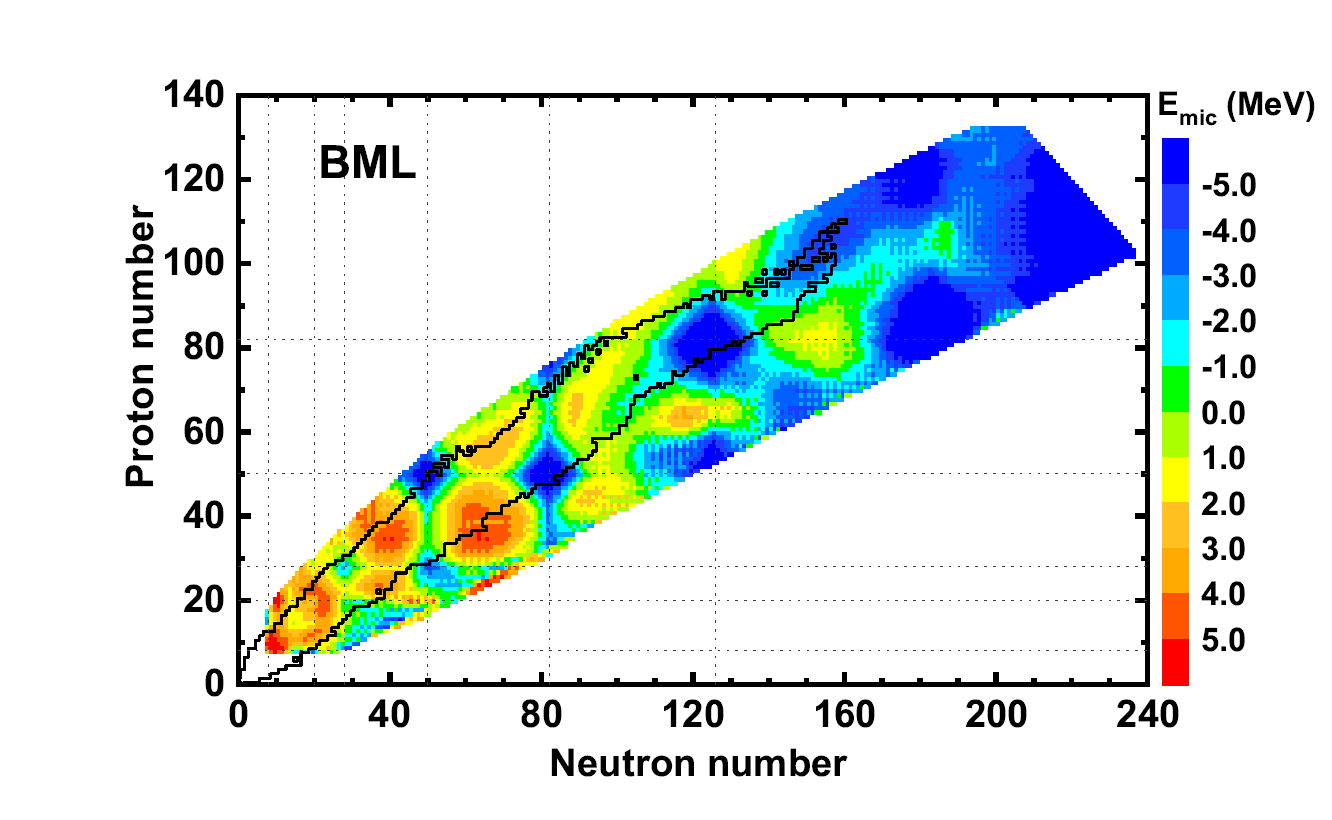}
\caption {(Color online) Microscopic correction energies $E_{\rm mic}$ of BML model. The contours show the boundary of nuclei with known masses in AME2020~\cite{WangM2021CPC} and the dotted lines denote the traditional magic numbers. The BML results are taken from Ref.~\cite{Niu2022PRC}.}
\label{Fig:EmicBML}
\end{figure}

The early applications of machine learning to nuclear mass predictions include those studies with ANN~\cite{Gazula1992NPA, Athanassopoulos2004NPA} and SVM~\cite{Clark2006IJMPB}. During the past decade, great progress has been achieved in nuclear mass predictions with machine learning methods. For the ANN, the Levenberg-Marquardt optimization algorithm~\cite{Zhang2017JPG}, the multiple training approach~\cite{Zhao2022NPAa}, and the multi-task ANN~\cite{Ming2022NST} has been developed to improve the mass predictions of existing nuclear models, which can get better accuracy and generalization ability of nuclear mass predictions. Apart from improving the accuracy of mass predictions, it is also found that the word-vectors from the shallow hidden layers of DNN are useful to predictions of other nuclear properties, e.g. $\alpha$-decay half-lives~\cite{Li2022PRC}. The BNN can avoid overfitting automatically by including prior distribution, quantify the uncertainties in its predictions, and evaluate the correlations among model parameters~\cite{Kejzlar2020JPG}, which was also employed to refine nuclear mass models~\cite{Utama2016PRC, Niu2019PRCb}. Based on the BNN, it is found that the inclusion of more physical features to the input layer of neural network, e.g. the quantities related to nuclear pairing and shell effects is very effective to further improve the performance of neural network in mass predictions~\cite{Niu2018PLB}, which illustrates the importance of the physical guidance to BNN predictions. By further improving the architecture of neural network, the Bayesian machine learning (BML) mass model was proposed recently, which crosses the accuracy threshold of the $100$ keV in the experimentally known region~\cite{Niu2022PRC}. The microscopic correction energies of BML model are shown in Fig.~\ref{Fig:EmicBML}, from which shell effects in nuclear properties can be revealed. Clearly, the BML model not only well reproduces the shell structures in the known region, but also predicts several remarkable structure features when extrapolated to the unknown region, e.g. the magic numbers around $Z = 120$ and $N = 184$ in the superheavy nuclei region. A recent development of the gradient boosting decision tree, the LightGBM was also used to refine nuclear mass models, which achieves very high accuracies for both training and testing data~\cite{Gao2021NST}. Moreover, the importance of input features in refining mass models was analyzed by investigating the correlation between the input characteristic quantities and the output, which may provide new insights for further developing nuclear mass models~\cite{Gao2021NST}. In addition, the RBF approach~\cite{Wang2011PRC} and its improved version of RBF with odd-even effects (RBFoe)~\cite{Niu2016PRC} have also been widely used to improve the predictions of nuclear masses~\cite{Niu2013PRCb, Zheng2014PRC, Ma2015JPG, Li2021CTP, Ma2017PRC, Niu2018SciB, Shi2019CPC}. By including a regularizer to reduce the risk of overfitting of RBF approach, the KRR~\cite{Wu2020PRC}, the KRR with odd-even effects~\cite{Wu2021PLB, Guo2022Symmetry}, and the gradient KRR (a multi-task learning framework)~\cite{Wu2022PLB} have been developed to improve the predictions of nuclear masses. Different from previous machine learning methods, the NBP method turns the predictions to classification problem, which also remarkably improves masses predictions of nuclear models by combining the $k$-means algorithm~\cite{Liu2021PRC}.

Nuclear charge radius is another important physical quantity of nuclear structure. With the development of laser spectroscopy \cite{YangXF1,YangXF2}, the measurements of nuclear charge radii have been extended to the regions far from $\beta$-stability line by using the optical isotope shifts, and about $1000$ known nuclear charge radii have been summarized and evaluated by combining the experimental data from various methods~\cite{Angeli2013ADNDT}. On the theoretical side, most microscopic and phenomenological nuclear models can reproduce the known data of charge radii with the rms deviation ranging from $0.07$ to $0.02$ fm. The machine learning has been used to directly describe nuclear charge radii with ANN~\cite{Wu2020PRC} or improve the theoretical predictions of charge radii with NBP~\cite{Ma2020PRC}, KRR~\cite{MaJQ2022CPC}, and BNN~\cite{Dong2022PRC} methods. It is found that the inclusion of the electric quadrupole transition strength and symmetry energy to the input layer of ANN plays an important role describing isotopic dependence of charge radii and the kinks of charge radii at the neutron magic numbers~\cite{Wu2020PRC}. After the refinements of KRR and BNN methods, the theoretical accuracies of charge radii are increased remarkably and the rms deviations to the known data is about $0.015$ fm.

Nuclear $\alpha$ and $\beta$ decay are basic properties of unstable nuclei, which play important roles not only in probing nuclear structure properties but also in studying astrophysical nucleosynthesis processes. The theoretical predictions of $\alpha$-decay half-lives have been remarkably improved with the RBF network~\cite{Ma2021CPC}. The Gaussian process provides a trustworthy method for predicting the $\alpha$-decay properties including energies and half-lives. Its predictions agree well with the experimental data and are also in good accordance with the theoretical results in the unknown region~\cite{Yuan2022CPC}. In addition, an unified description of $\alpha$-decay and cluster radioactivity half-lives has been achieved with the accuracy better than the current physical model~\cite{Zhao2022JPG}. For $\beta$ decay, the half-lives~\cite{Costiris2009PRC, SSPMA-2021-0299} and $\beta$-delayed one-neutron emission probabilities~\cite{Wu2021PRC} have been studied by the ANN. By carefully designing the input and output layers, the BNN can introduce the known physics described by the Fermi theory of $\beta$ decay, and the dependence of half-lives on pairing correlations and decay energies. Table~\ref{tab:tmsT12} shows the rms deviations $\sigma_{\rm rms}(\log_{10} T_{1/2})$ of BNN predictions~\cite{Niu2019PRCa} with respect to the experiential $\beta$-decay half-lives~\cite{Kondev2021CPC} in comparison with those of the quasiparticle random phase approximation based on the relativistic Hartree-Bogoliubov (RHB+QRPA)~\cite{Marketin2016PRC}, Skyrme-Hartree-Fock-Bogoliubov (SHFB+QRPA)~\cite{Minato2022PRC},  finite-range droplet model (FRDM+QRPA)~\cite{Moller2019ADNDT}, and the finite-amplitude method based on the SHFB (SHFB+FAM)~\cite{Ney2020PRC}. It is clear that the BNN can reproduce the experimental half-lives with a high accuracy, which is much better than those other nuclear models~\cite{Niu2019PRCa}. Nuclear $\beta$-decay half-life predictions can also be improved by learning those not-well-determined parameters of nuclear models with neural network, e.g. the isoscalar pairing strengths in the QRPA calculations~\cite{Minato2022PRC}. It is found that the extrapolations of this method can keep reasonable isotopic trends and only have moderate uncertainties.

\begin{table}[htbp]
\caption {The rms deviations $\sigma_{\rm rms}(\log_{10} T_{1/2})$ of $\beta$-decay half-lives with respect to the experiential data~\cite{Kondev2021CPC} for various models. The values denoted by $T_{1/2}^{S6}$, $T_{1/2}^{S3}$, and $T_{1/2}^{S0}$ are the rms deviations with respect to the experimental data of three nuclear sets with $T_{1/2} \leqslant 10^6$ s, $T_{1/2} \leqslant 10^3$ s, and $T_{1/2} \leqslant 1$ s, respectively.}
\begin{center}
\begin{tabular}{cccc}
\hline
\hline
Method        & $T_{1/2}^{S6}$     & $T_{1/2}^{S3}$    & $T_{1/2}^{S0}$ \\		
\hline
BNN~~~~~        &0.420~~~~~  &0.350~~~~~  &0.225  \\
RHB+QRPA~~~~~   &1.025~~~~~  &0.905~~~~~  &0.461  \\
SHFB+QRPA~~~~~  &0.837~~~~~  &0.684~~~~~  &0.469  \\
FRDM+QRPA~~~~~  &0.789~~~~~  &0.596~~~~~  &0.390  \\
SHFB+FAM~~~~~   &0.811~~~~~  &0.710~~~~~  &0.397  \\
\hline
\hline
\end{tabular}
\end{center}
\label{tab:tmsT12}
\end{table}

The machine learning can also help us to describe nuclear excited properties, e.g. nuclear low-lying excitation spectra and giant dipole resonances (GDR) parameters. By including an input related to nuclear collectivity besides proton and neutron numbers, the BNN can well describe nuclear low-lying excitation energies in a large energy scale from about $0.1$ MeV to about several MeV, whose accuracy is significantly better than the sophisticated microscopic collective Hamiltonian model~\cite{Wang2022PLB}. The inputs of the BNN can be carefully selected from many ground-state properties guided by Pearson's correlation coefficients between them and the output GDR energies, which effectively reduce the predicted errors and avoid the overfitting~\cite{WangXH2021PRC}. To simultaneously describe multiple physical quantities with a single neural network, the multi-task ANN was developed to describe both GDR energies and widths~\cite{Bai2021PLB} or multiple low-lying excitation energies~\cite{Wang2022NPR}. The accuracies of multi-task ANN are much better than traditional nuclear models, which proofs the feasibility of studying multi-output nuclear physical problems with multi-task ANN.

The machine learning has also been employed to study other nuclear properties with enough experimental data, e.g. nuclear spins and parities~\cite{Gernoth1993PLB} as well as nuclear magnetic moments~\cite{Yuan2021CPC}. Apart from learning the experimental data, the machine learning can also be used as emulators of nuclear models by learning model predictions, e.g. a KRR emulator for the kinetic energy part of the energy density functional~\cite{Wu2022PRC,Ren_2021}, DNN emulators for the density distributions calculated by Skyrme density functional theories~\cite{Yang2021PLB}, DNN emulators for the constrained Hartree-Fock-Bogoliubov calculations~\cite{Lasseri2020PRL}, and ANN extrapolation tools for $ab~initio$ calculations to very large model space~\cite{Negoita2019PRC, Jiang2019PRC}.

\subsubsection{Fission yields}

Nuclear fission is one of the most complex processes in nuclear theory and has strong application motivations. 
Current phenomenological models can well describe the fission observables where abundant experimental data are available,
but they are questionable when being extrapolated to unknown regions.  
Microscopic fission theory can in principle self-consistently describe multiple fission observables but
its accuracy needs to be improved. ML provides an alternative method to evaluate the
nuclear fission data, in particularly for treating the noisy, discrepant and incomplete data. 
ML can take into account underlying data correlations to infer. 

Fission product yields (FPY) are the key fission observables being correlated with other observables. In major nuclear 
data libraries, such as ENDF, JEFF, JENDL, and CENDL, the evaluated fission yields are only available
at thermal neutron energies, 0.5 MeV and 14 MeV.  The energy dependence of FPY is ascribed
to energy dependent shell effects, dissipation effects, and
prompt neutron emissions.
The evaluation of fission yields at other energies
for fast reactors is very desirable. In experiments, it is very difficult to measure the yields of
full fragments. In this case, the Bayesian neural network is applied to learn existing libraries of
evaluated fission yields as well as incomplete experimental data points to infer the full
fission yields~\cite{WangZA2019}. The input of the network is given by ($Z_i$, $N_i$, $A_i$, $E_i$), which includes  
the charge number $Z_i$ and neutron number $N_i$ of the fission nuclei, the mass number $A_i$ of fragments, and 
the excitation energies $E_i$ of the fission nuclei. 
 The results can reasonably reflect the uncertainties. The uncertainties 
become larger when neighboring data points are sparse. The evaluated fission yields can reasonably 
show the energy dependence with increasing fission yields at the symmetric fission channel. 

\begin{figure}[htbp]
\centering
\includegraphics[scale=0.21]{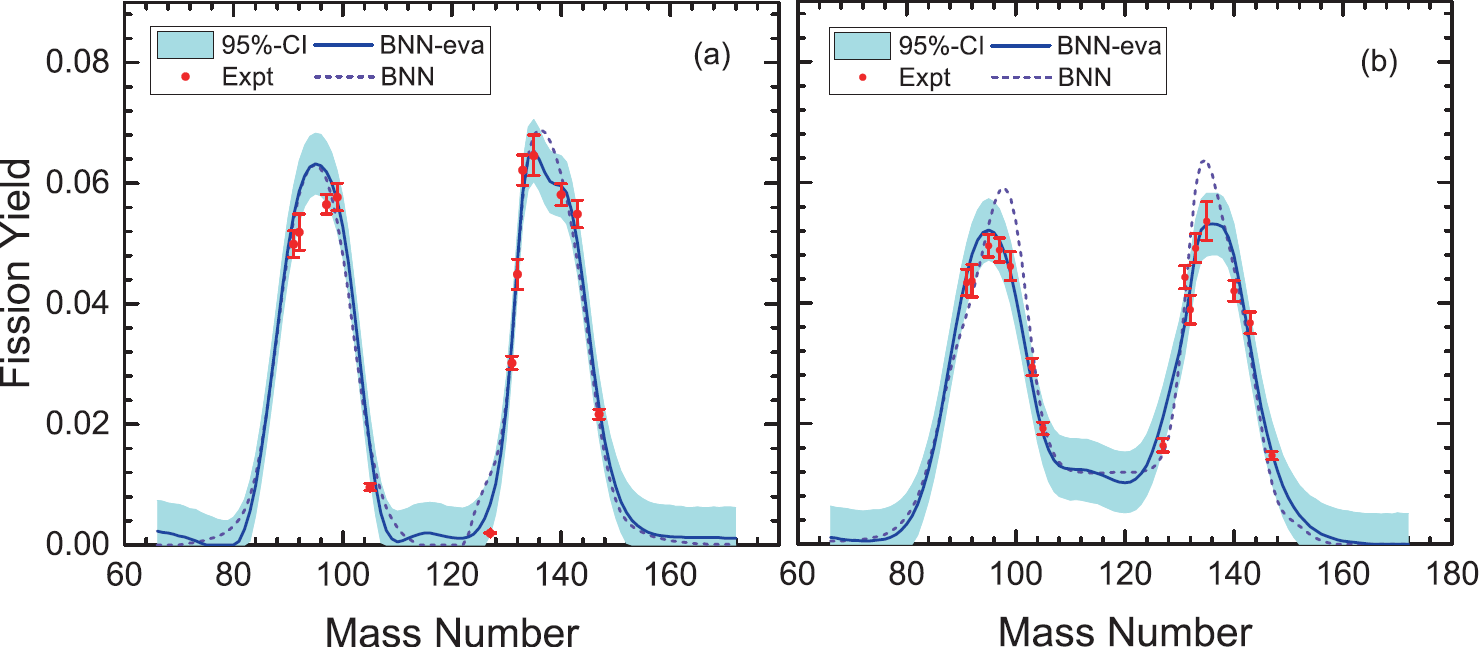}
\caption{(Color online) The BNN evaluation of fission yields of n+$^{235}$U at energies of 1.37 MeV (a) and 14.8 MeV (b).
  The dashed line and solid lines denote the BNN prediction without and with learning the experimental data.
  The shadow region corresponds to the confidence interval estimated at 95$\%$.  The figure is taken from Ref.~\cite{WangZA2019}.    
}
\label{fig:fission1}
\end{figure}

In addition to the mass yields, the fission charge yields  have also been studied~\cite{QiaoCY2021}. In two experiments, 
the fission charge yields from the compound nucleus $^{239}$U have large discrepancies. 
It was reported that the charge yields around Sn and Mo are exceptionally small in the $^{238}$U(n, f) reaction  but it
is normal in a later experiment on $^{239}$U via transfer
reactions. To evaluate the charge yields, the odd-even effects are included in the input data. 
The BNN evaluations don’t obtain
abnormal small charge yields around Sn and Mo isotopes
as reported  and are consistent with the latest
experiment. 
BNN can be very powerful for evaluations of fission data but its potential has not been 
fully realized. 
To this end, the multi-layer BNN has been tested~\cite{WangZA2021}. With similar number of parameters,
the deeper networks don't demonstrate advantages. On the other hand, the shallow networks involve much 
information of input data and can be well trained. 

Recently Bayesian data fusion has been applied to evaluate the full two-dimensional fission yields~\cite{WangZA2022}. 
Data fusion refers to the process that information from individual datasets sharing at least a number
of variables is merged. Data fusion is expected
to produce more consistent, accurate and useful information than separated data sources,
by including non-local and
high-dimensional correlations, in analogy to long-range and many-body interactions in quantum systems. 
The inference would be less
precise when data in some energies is sparse, however, its
correlations with heterogeneous data in other energies can be useful to improve the inference. 
It is demonstrated that  the evaluation of extremely incomplete
independent fission yields in terms of energy dependencies, namely a large gap between 2 and 14 MeV, 
can profit from the data fusion
of a more comprehensive coverage of cumulative fission
yields~\cite{WangZA2022}. For the data fusion, it is crucial to take into account the experimental
uncertainties by the likelihood function. The results show that the uncertainty propagation is a comprehensive effect and would not significantly change due to a
few specific data, due to complex uncertainty correlations~\cite{Yijiayi}.
It is expected that the Bayesian data fusion can facilitate the maximum utilization of imperfect raw nuclear data. 
The BNN evaluations are comparable to
GEF evaluations for particular evaluations. 
Besides, the mixture density network has been applied to learn parameters of Gaussian functions to simulate fission yields~\cite{Lovell2020}.
Recently a tensor decomposition model has also been used to evaluate fission yields by reconstructing evaluated database~\cite{SongQF2022}. 

\subsection{Nuclear reaction}

Nuclear reaction describes how two nuclei react with each other at certain conditions, which is an important field in nuclear technology and engineering, nuclear safety and medicines, as well as nuclear physics. Especially for low and intermediate energy nuclear physics, the nuclear reaction can also be used to investigate the properties of compressed and excited nuclear matter as well as to map out the initial structure of nuclei and understand the nuclear potential based on the accumulated big data.  Thus, constructing a predictable model for describing and inferring the interesting nuclear reaction data will be important and a ML algorithm will be a good choice.



\subsubsection{Refinement of the description of reaction data}


Spallation reactions are nuclear reactions where a light projectile bombards a heavy target nucleus at the energy up to GeV. removing one or more nucleons from target and producing lighter nuclei. Such reactions occur during the propagation of cosmic rays in the galaxy's interstellar medium. Spallation of abundant CNO nuclei is the main nucleosynthesis mechanism producing the light and fragile nuclides of Li, Be, and B, which are destroyed by thermonuclear reactions in stars. In addition, spallation reactions play an important role in neutron sources, nuclear waste disposal, radiation protection, rare isotope production, nuclear and particle physics experiments~\cite{Tang,Su}.

However, spallation reactions are complex due to the wide range of incident energies and abundant fragments involved.
There are three main mechanisms in spallation reactions, evaporation, fission, and multifragmentation. To describe the spallation reactions, several Monte Carlo intra-nuclear cascade model, transport models, and statistical decay models are employed. Due to the large range of reaction energy and big final state space. The predictions by model calculation still have non-ignorable bias comparing experimental measurements. 
ML is proved to be a powerful correction tool for traditional models.

A physical model dependent BNN, was developed for a more accurate evaluation of production cross sections \cite{Song2022CPC}. The BNN algorithm is used to improve the prediction accuracy after learning the residual error between experimental data and calculations by the IQMD-GEMINI++ model. The deviation between calculations and experimental data were reduced to within 0.4 order of magnitude.

The BNN + sEPAX methods are applied to construct predictive models for fragment cross sections in proton-induced nuclear spallation reactions \cite{Peng2022JPG}.  It is found that the BNN + sEPAX model performs good extrapolations based on less information due to the physical guidance of the sEPAX formulas. 

To describe nuclear reactions for incident particles including photon, neutron, proton, deuteron, triton, $^3$He, and $\alpha$ particles within the energy range of 1 keV to 200 MeV, a TALYS code is widely used~\cite{KONING20122841}. Recently, an iterative Bayesian Monte Carlo (iBMC) method is employed to find the optimal model and parameter sets of the TALYS code system~\cite{Alhassan}. The Maxwellian-averaged neutron-capture cross-sections predicted by the TALYS code can be further refined by the KRR method~\cite{Huang_2022}. It is found that the rms of the relative errors between the experimental data and the TALYS predictions is significantly reduced with the KRR approach. The inclusion of cross-section data with different temperatures is helpful to further improve the predictive performance of the KRR approach.

\subsubsection{Probing the initial state of nuclei and reaction geometry}

To understand the nuclear collision process, the initial state of nuclei should be modeled or quantified precisely. Mapping the complex final states to the initial state features are challenging tasks. The universal approximate ability of the deep neural networks shows great advantages in mapping the high dimensional features and correlations in the final states to the initial states.

Clustering is an exotic phenomenon in nuclei, especially in light nuclei. The clustering structure could make the final state particles anisotropic distributed \cite{Ma_book,Shi}. In the $^{12}C$ / $^{16}O$ + $^{197}$Au collisions at relativistic energies, the azimuthal angle and transverse momentum distributions of charged pions are used to retrieve the cluster structures of $^{12}C$ / $^{16}O$\cite{He2021PRC}. A Bayesian convolutional neural network is employed. The neural network model can distinguish the clustering and non-clustering initial states within 95\% accuracy. For the input data set, thousands final states are merged to one histogram to reduce the fluctuation. 

Deformation is a ubiquitous feature of most atomic nuclei and it is difficult to disentangle in heavy-ion collisions due to the highly complex and dynamical nature of the collisions. Deep convolutional neural networks are trained to predict two deformation parameters $\beta_2$ and $\beta_4$ from physical observables obtained through theoretical simulations. The magnitude of the nuclear deformation from event-by-event correlation between the momentum anisotropy or elliptic flow and total number of charged hadrons are successfully extracted. A Regression Attention Mask algorithm is designed to interpret what has been learned by the neural network \cite{Pang2019aqb}. 
Many experiments and theories have shown that the quadrupole deformation of the ground state of the nuclei will affect the observed quantities of the final state of the heavy ion reaction. However, the initial deformation signal will be gradually masked with the dynamic evolution during the reaction. In Ref.~\cite{gao2022deformation}, the three-particle spectrum distributions in the final state of the reaction given by the presence or absence of deformation effects in the training initialization process of the LightGBM algorithm are used to explore the ability of identifying the nuclear deformation in the particle distribution of the final state. And the window which is sensitive to the initial deformation in the final state particle spectral distribution is given through the analysis of the importance feature.

The impact parameter is another important reaction geometry for heavy-ion collisions (HICs), which describes the distance between the centers of the two colliding nuclei in a classical view and influences the reaction mechanism and final observables. Thus, the knowledge of the impact parameter or centrality of HICs is crucial.
However, it cannot be measured directly in experiments. The traditional methods estimate the impact parameter from multiplicity of final state charged particles and transverse momentum of light particles under the assumption of averaged multiplicity of charged particles decreases monotonically when the impact parameter increases. In the 1990's, mapping impact parameter from the final observables by using simple neural network were proposed in Refs.~\cite{Bass1996PRC,Haddad1997PRC}. 
%

With the help of the developments of ML algorithms, many new efforts on estimation of the impact parameters of HICs have been devoted with supervised machine learning. In high energy heavy-ion collisions, ANN (or named Multilayer Perceptron, MLP) and CNN are applied to mapping the charged particle energy spectrum to the impact parameters\cite{Xiang2022CPC}, and a better accuracy can be obtained with certain algorithms. In Refs.~\cite{Li2021PRC, Li_2020JPG}, reconstructing impact parameters from the HIC observables were investigated by using CNN and LightGBM. The datasets they used are obtained from the ultrarelativistic quantum molecular dynamics (UrQMD) model simulations. It is found that the obtained mean absolute error $\Delta b$ increases with the decreasing of beam energy. For Sn + Sn at 270 MeV/nucleon, the $\Delta b$ reach to 0.8 fm~\cite{Li2021PRC}. With the datasets generated by CoMD model, a CNN algorithm for impact parameter determination is extended to lower energy region, from several tens to one hundred MeV/nucleon~\cite{Zhang2022PRC}. 

However, the generalizability, the model dependence of ML and the inherent fluctuation of observables with respect to impact parameter~\cite{Lili2022SCPMA} affect the accuracy of predicting unknown data or extracting the exact value of impact parameter. 
In Ref.~\cite{Lili2022arXiv}, a model-independent method which also considers the inherent fluctuation mechanism was proposed for reconstructing the impact parameter distributions from multi-observables. 
This method combines a Bayesian method to unsupervised ML algorithms, i.e., k-means algorithm, and can be used to multi-observables in experiments.
\subsubsection{Reaction mechanism and phase transition}

For low energy nuclear reactions, the optical model potential (OMP) is the most successful macroscopic model in the field of nuclear reaction. It describes the nuclear interactions phenomenologically, analogous to the refraction and absorption of a light wave by a medium with complex refractive index in optics. Due to the causal relationship, the outgoing wave cannot be generated before the arrival of the incident wave, which leads to the named dispersion relation between the real and imaginary parts of the OMP.
The validity of the dispersion relation in exotic systems is controversial which may be a common phenomenon for exotic nuclear systems. The transfer reaction method provides a strong evidence that the dispersion relation does not hold for $^6$He+$^{209}$Bi at energies around the Coulomb barrier. However, according to the analysis by using the frequentist approach of bootstrap method, the dispersion relation exists for $^6$Li+$^{209}$Bi and $^{6}$He+$^{208}$Pb. This approach is unsatisfactory especially for the unrepeatable situations and for the limited sequence of repetitions. The Bayesian method considers that the probability is a measure of uncertainty of our knowledge of the physical world. Thus the probability of model or parameters can be assigned, which is thought more reliable and flexible than the frequentist approach in quantifying uncertainties on the reaction observables. 

\begin{figure}[htbp]
\centering
\includegraphics[scale=0.3]{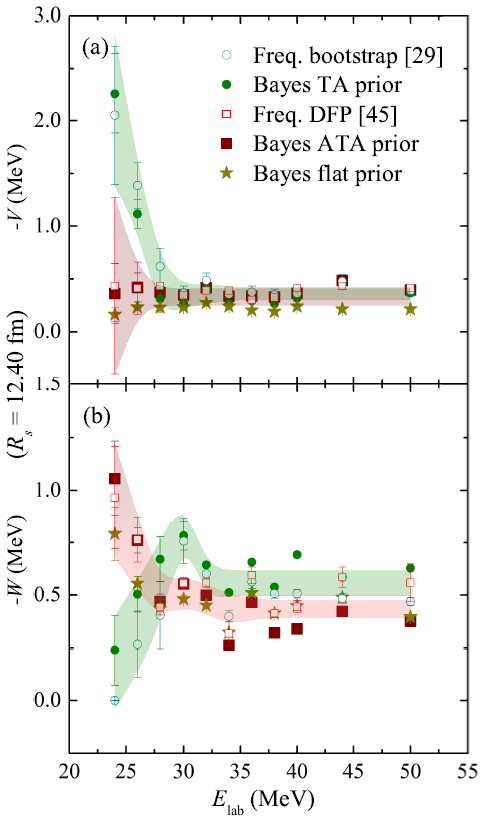}
\caption{(Color online) Bayesian inference on the energy dependence of the real (a) and imaginary (b) potentials at the sensitivity radius of 12.40 fm for the $^6$Li+$^{209}$Bi system. Full squares, circles, and stars denote the Bayesian results with ATA prior, TA prior and flat prior respectively. The figure and references [29] and [45] in panel (a) are taken from Ref.~\cite{Yang2020PLB}.    }
\label{fig:OMP-Li6Bi208}
\end{figure}
In Ref.~\cite{Yang2020PLB}, the Bayesian inference has been implemented to study the properties of the reaction systems with exotic nuclei, for instance, the gross feature of the phenomenological interaction potential. As shown in Figure \ref{fig:OMP-Li6Bi208}, the results of Bayesian statistics strongly depend on the imposed prior distributions. Therefore, the Bayesian method has to be used with extreme caution, since the improper prior knowledge could lead to completely wrong conclusions which can be avoided with accumulating more data in future.

For low-intermediate energy HICs, multifragmentation is an important mechanism since it is also related to the liquid-gas phase transition of finite nuclear system. 
Over the past several decades, the nuclear liquid-gas phase transition has been studied based on the heavy-ion collisions at intermediate and relativistic energies \cite{Rev1,Ma_PRC,Ma_PRL,Deng,Liu} and hadron-nucleus collisions at relativistic energies \cite{Isis}.

\begin{figure}[htbp]
\centering
\includegraphics[scale=0.28]{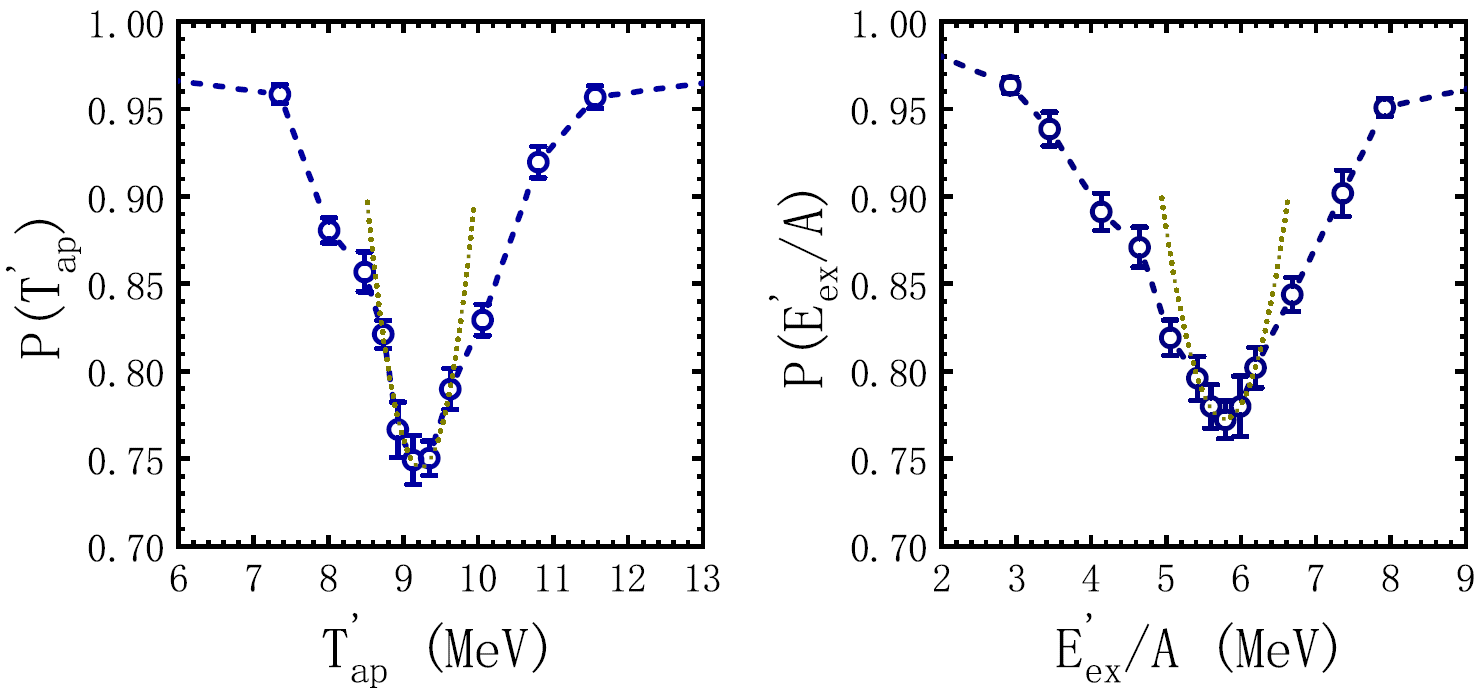}
\caption {(Color online) \small The performance curve $P(T_{\rm ap}')$ and $P(E_{\rm ex}'/A)$, i.e., the testing accuracy as a function of the proposed temperature $T_{\rm ap}'$ and transition excitation energy $E_{\rm ex}'$, respectively. The yellow dotted lines represent a parabolic fit of the lowest five data points with errors \cite{Wang2020PRR}.}
\label{Fig:PC}
\end{figure}

In Ref.~\cite{Wang2020PRR}, the autoencoder method is adopted to study the nuclear liquid-gas phase transition. The encoder encodes the input event-by-event charge-weighted charge multiplicity distribution of quasiprojectile fragments to a latent variable, the decoder part tries to decode the latent variable back to the original distribution. The averaged latent variable $P(T'_{ap})$ as a function of apparent temperature $T'_{ap}$ or the excitation energy per nucleon $E_{ex}/A$ exhibits a sigmoid pattern, which indicates the trained autoencoder network treats the low- and high-temperature regions differently. The autoencoder network is capable of classifying different phases of nuclei directly from the final-state information of the heavy-ion experiment. With the confusion scheme, the neural network is trained with data that are deliberately labeled incorrectly according to a proposed critical point, and the phase transition properties can be deduced from the performance curve. A $V$ shape performance curve shows the nuclear liquid-gas phase transition. In order to properly include the uncertainty of the obtained limiting temperature, a BNN is adopted to capture the uncertainty.

In Ref.~\cite{Song2021PLB}, a DNN is trained to determine the apparent temperature $T'_{ap}$ of HICs at intermediate-to-low energies through their final-state charge multiplicity distribution $M_c(Z_{cf})$. Based on the IQMD simulation datasets, the relation between the final-state $M_c(Z_{cf})$ of a nuclear source and its corresponding temperature is established through training a DNN. The trained DNN can predict the apparent temperature within an error of 0.62 MeV, which is small enough for applying it to analyze the reaction dynamics. With the temperature determined by DNN, the caloric curve of the reaction is studied and the apparent temperature of nuclear liquid-gas phase transition can be determined.

\subsection{Nuclear matter}

The isospin asymmetric nuclear equation of state (EOS) describes the relationships between the energy,
pressure, temperature, density and isospin asymmetry. Within the parabolic approximation, the energy per nucleon of nuclear system is described as,
\begin{equation}
	\label{eos}
	E(\rho,\delta) = E(\rho,\delta=0)+S(\rho) \delta^2+\cdots. 
\end{equation}
$\rho_n$ and $\rho_p$ are the neutron and proton densities, respectively, and $\delta = (\rho_n-\rho_p)/(\rho_n+\rho_p)$. In theory, large uncertainties of density dependence of the symmetry energy away from normal nuclear matter are predicted and it stimulates lot of efforts to constrain isospin asymmetric nuclear equation of state, especially the density dependence of symmetry energy $S(\rho)$ in past 30 years \cite{Tsang,Li22,Liu21}. Generally, constraints of symmetry energy can be obtained from the investigation of properties of nuclear structure, heavy ion collision and neutron stars. Recent review papers \cite{Steiner,LiBA} are suggested for more details.

With the progress of study of the density dependence of the
symmetry energy, tight constraints become urgently requisite,
which need the efforts from experimental measurements, improvements
of transport models, and understanding the physics parameter
correlations. For example, we know that there are not only $S_0$
and $L$ in the Taylor expansion of $S(\rho)$,
\begin{equation}
    S(\rho)=S(\rho_0)+L(\frac{\rho-\rho_0}{3\rho_0})+\frac{K_{sym}}{2}(\frac{\rho-\rho_0}{3\rho_0})^2+\frac{Q_{sym}}{6}(\frac{\rho-\rho_0}{3\rho_0})^3
\end{equation}
but also the high order terms, such as curvature $K_{sym}$ and
skewness $Q_{sym}$ of $S(\rho)$ and so on. Or, the symmetry energy will not only depend on $S_0$ and $L$, but also on incompressibility $K_0$, isoscalar effective mass $m_s^*$, and isovector effective mass $m_v^*$~\cite{ZYX20}. Obviously, the uncertainties
of $K_{sym}$ and $Q_{sym}$ can influence the constraints on
$S_0$ and $L$, or the density dependence of the symmetry
energy. Thus, low-biased analysis on multi-dimensional parameter space are needed and which naturally requires some machine learning algorithms since it can avoid the curse of dimensionality in certain level.

\subsubsection{EOS and symmetry energy from nuclear structure}

For constraining the isospin asymmetric EOS, the usually used observables in nuclear structure studies are the neutron skin, electric dipole polarizability, isovector giant dipole resonance (IVGDR) and isoscalar giant quadrupole resonance (ISGQR), binding energy, charge radius and breathing mode energy. Among them, the neutron skin thickness $\Delta r_{np}$ \cite{PRL1,Hu,Wei,MaCW} 
in heavy nuclei was known as one of the most sensitive terrestrial probes of the symmetry energy around 2/3 of the saturation density, i.e., $2/3\rho_0$. In Ref.\cite{Xujun2020PRC}, a Bayesian inference method is used to obtain the probability distribution function of the slope of symmetry energy $L$ with the available neutron skin data for $^{116,118,120,122,124,130,132}$Sn from hadronic probes. The neutron skin data for Sn isotopes gives $L = 45.5^{+26.5}_{-21.6}$ MeV surrounding its mean value or $L = 53.4^{+18.6}_{-29.5}$ MeV surrounding its maximum a posteriori value, respectively. By using the recent published results on the neutron skin of $^{48}$Ca and $^{208}$Pb by the CREX and PREX-II, an effort on extraction of symmetry energy has also been performed by using Bayesian inference method~\cite{ZhenZhang2022arXiv}. Their results show the symmetry energy and neutron skin of $^{48}$Ca and $^{208}$Pb are compatible with each other at 90\% confidence level. In addition, a Bayesian analysis of the electric dipole polarizability, the constrained energy of isovector giant dipole resonance, the peak energy of isoscalar giant quadrupole resonance, and the constrained energy of isoscalar giant monopole resonance in $^{208}$Pb are also performed\cite{ZhenZhang2021CPC}, and they got the probability distribution function of isoscalar and isovector effective masses in nuclear matter at saturation density. The results show a positive isospin splitting of nucleon effective mass in asymmetric nuclear matter of isospin asymmetry.


\subsubsection{EOS and symmetry energy from HICs}

For low-intermediate energy HICs, there is fluctuation mechanism that comes from the microscopic statistical fluctuation of initial states and random scattering during the reaction process. This inherent fluctuation leads to the observables distributing in a wide range even for given EOS parameters. 
Thus, the probability to be a measure of uncertainty regarding our knowledge of the physical world has to be considered. To describe the distribution of constrained parameters, a Bayesian inference method is suitable.

In Ref. \cite{MORFOUACE2019PLB}, ImQMD model is used to constrain the symmetry energy related parameters in four-dimensional parameter space. They are symmetry energy coefficient $S_0$, the slope of symmetry energy $L$, isoscalar effective mass $m_s^*$, and isovector effective mass $m_v^*$. By using the Bayesian inference, the constraints of the symmetry energy in this four-dimensional parameter space with single ratios of neutron and proton spectra in central $^{112}$Sn + $^{112}$Sn and $^{124}$Sn + $^{124}$Sn collisions at 120 MeV/u is obtained. The distributions of $S_0$, $L$, $m_s^*$, and $m_v^*$ are obtained, which can be used to learn the correlation between parameter pairs.

In Ref.~\cite{FPLi2020NPR} the CNN and LightGBM methods were employed to identify the nuclear EOS with $K_0=200$ and $380$ MeV by analyzing the proton spectra in transverse momentum and rapidity at the final state. It is found that the classification accuracy for event-by-event can reach 85\%, while that from event-summed input can increase to 95\%. Furthermore, using the Prediction Different Analysis method can help us obtain the sensitive region of proton spectra. 
Due to the fluctuation mechanism, accurate constraint of EOS parameter is impossible and their distributions are expected by the machine learning algorithm.

In Ref.~\cite{Wang2021PLB}, a deep CNN was developed to decode signatures of the nuclear symmetry energy in HICs by learning the transverse momentum and rapidity distributions of nucleons. In the calculations, five Skyrme types of symmetry energy are used. They are Skz4, SLy230a, SV-sym34, SkI2 and SkI1, and the corresponding slope of symmetry energy $L$ ranges from 5.8 to 159.0 MeV. In inference the symmetry energy, both classification and regression tasks are performed with labeled data-set from UrQMD model simulation. It is found that the trained CNN is able to identify the fingerprints of the nuclear symmetry energy, which is a challenge to conventional methods because the effects of nuclear symmetry energy  might be easily washed out by strong fluctuations. In the latest attemp~\cite{WangYJ2022PLB}, the nuclear symmetry energy was further decoded from a large set of observables in HICs on an event-by-event basis by the LightGBM method, and its slope parameter $L$ can be predicted but with a mean absolute error of approximately 30 MeV.

\subsubsection{EOS and symmetry energy from neutron star}

A neutron star is the remnant of a supernova explosion of a massive star, and its interior contains the densest
nuclear matter in the universe. Roughly, neutron stars have surface, outer crust, inner crust, outer core an inner core. From densities ranging from saturation density ($\sim\rho_0$) to $\sim 3\rho_0$ of inner core, it was thought that it mainly consists of neutron, proton and leptons. The properties of neutron star, such as mass-radius relation, tidal deformability, maximum mass, are closely related to the nuclear equation of state, especially the isospin asymmetric nuclear matter at high density. As mentioned before, the symmetry energy of isospin asymmetric nuclear equation of state not only depends on the $S_0$ and $L$, but also the high order terms of expansion coefficients in the density region up to 3$\rho_0$.

In Ref.~\cite{Xie2020ApJ}, a Bayesian method has been used to infer the posterior probability distribution functions of the high-density nuclear symmetry energy. The data they used is the radii of canonical neutron stars~\cite{Xie2019ApJ} and more massive neutron stars~\cite{Xie2020ApJ}, such as the radii of canonical neutron stars (NSs) reported by the LIGO/VIRGO and NICER Collaborations based on their observations of GW170817 and PSR J0030+0451. Based on these results, how future radius measurements
of more massive NSs will improve our current knowledge about the EOS of superdense neutron-rich nuclear
matter are also analyzed.

\subsection{Experiments}
Machine learning benefits the nuclear experiments in many ways, including track reconstruction, trigger efficiency, event identification, complex system control, and firmware-base applications.

\subsubsection{Event identification and reconstruction}

Event identification and background reduction are the most important tasks in neutrino experiments, dark matter experiments, and neutrinoless double beta decay experiments due to their rare event rate as well as the relatively high background.
The convolutional autoencoder machine learning technique is employed to reduce the background in Daya Bay experiment \cite{Kohn_2017}.
In JUNO experiment, to achieve the neutrino mass ordering determination, the tasks of the vertex and energy reconstructions are important and challenging. Several machine learning approaches  are employed, including BDT, DNN, ResNet, VGG, and GNN. The models are trained on Monte Carlo simulation data. The vertex coordinates is around $\sigma_{x,y,z}$ = 10 cm at $E_{vis}$ = 1 MeV and decreases at higher energies, and the energy resolution is around $\sigma_E$ = 3\% at $E_{vis}$ = 1 MeV, which resolution satisfies the requirements posed by the JUNO experiment \cite{Li2022,QIAN2021165527}.
In PandaX experiment \cite{Pandax-2}, the performance of CNN in double beta decay events classification is much better than topological method \cite{Qiao2018}.
In the EXO-200 experiment, the deep neural networks are applied to the data analysis. The total energy and position are directly reconstructed from raw digitized waveform \cite{Delaquis_2018}.
In the MicroBooNE experiment, the convolutional neural networks are employed for the particle identification, event detection \cite{Acciarri_2017}.

The Active-Target Time Projection Chamber (ATTPC) is an important detector due to the capable of full three-dimensional reconstruction of charged particles, especially for the low energy nuclear experiments, as well as the gas target makes the detection of low energy decay available. Due to the short track, the tracks fitting and classification are both more difficult than the case in high energy nuclear experiments. In the ATTPC detector at the National Superconducting Cyclotron Laboratory (NSCL), the CNN methods are employed capture the data from the simulations and experimental data \cite{KUCHERA2019156}. 

For the electromagnetic calorimeters, the energy resolution and spatial resolution are key features. Deep learning method can benefit both the energy resolution \cite{Simkina2825519} and the spatial resolution \cite{Wang_2019JOI}.

\subsubsection{Complex system control and FPGAs applications}

ML is widely used in the control of complex equipment. In the Hefei Light Source II (HLS-II) storage ring, machine leading method is used to improve the feedback accuracy of the tune feedback system, and the betatron tune stability \cite{Yu2022NST}. 
Firmware base ML applications are promising in improving the performance of data acquisitions. In  the DUNE experiment, a hardware-accelerated Deep Neural Networks method is proposed for real-time data processing and data selection \cite{Jwa8909784}. An general Python package for machine learning inference in Field Programmable Gate Arrays FPGAs is developed. With this package, the traditional machine learning models can be easily translated into HLS that can be configured into FPGAs, for example the BDT, RNN, and CNN \cite{Duarte2018ite}

\subsection{Other applications}
The interdisciplinary research of ML and physics is developing rapidly, and new directions are constantly emerging. As an example, how neural network combines with theoretical calculations to be a new ab-initio computing strategy of quantum many body system is explained. Some ML applications in nuclear technologies and nuclear applications are also introduced.

\subsubsection{Ab-initio calculations of nuclei}

$Ab~initio$ nuclear calculations are based on realistic nuclear forces and in principle are the most accurate theoretical methods by including as many as possible many-body correlations. However, the computing costs  grow explosively towards heavier nuclei. 
For $ab~initio$ calculations of nuclei, a possible machine learning based method is developed from the variational Monte Carlo (VMC), which uses variational principle and Monte Carlo sampling to obtain the best parametrized trial wave-function \cite{Hermann2022arXiv}. Some calculations of certain quantum many-body problems for Bosonic and lattice systems are reported \cite{Saito2018}. By carefully designing the neural network to satisfy the anti-symmetric correlation of fermions, many electron systems  can be calculated \cite{HAN2019108929}.
Compared to the molecules, the interactions among the many-nucleon system is more complicated than the pure Coulomb interaction. With simple assumptions of nuclear potential, the neural network based VMC method only works for the very small nuclei \cite{KEEBLE2020135743}. For larger nuclei, the high-order corrections have to be included. By involved a leading-order pionless effective field theory hamiltonian, the nuclei with up to $A$ = 4 nucleons are calculated successfully \cite{Adams2021PRL}. Coupling NN and spherical harmonics, the nuclei with up to $A$ = 6 nucleons can be calculated \cite{Gnech2021wfn}. As the hidden fermionic degrees of freedom is involved in the variational wave functions, the number of variational parameters can be reduced relatively. Then the ground state wave function of $^{16}O$ can be calculated \cite{Lovato2022PhysRevResearch}. The neural network based nuclear many-body problem calculations are still at the very beginning. The biggest challenge is how to treat the complicated interactions between nucleons. But the neural network based methods and other machine learning based methods are still very attractive.

\subsubsection{Nuclear technology and applications}

Using the event mode sequence information of radionuclides, a well trained sequential Bayesian method can achieve the rapid identification of radioactive substances \cite{Li2021}.
The machine learning method, e.g. SVM \cite{Arahmane2021} and neural networks \cite{Zuo2021NST}, are employed to discriminate neutrons from gamma rays of signals in plastic scintillator neutron detectors. 
A machine learning method is employed to automatically identify and analyze the TEM images of helium bubbles \cite{Wu2021NST}. The helium bubble images are used as model inputs. The clustering analysis is used for better helium bubble identification. The position is evaluated by the Gaussian mixture analysis.


\section{Summary and outlook}

In this review paper, we have summarized the frequently used machine learning algorithms and applications for emphasizing the goal and significant role of ML. Different from traditional numerical tools, the goal of ML is to infer from data rather than fitting or simulation, which is explained by the introduction of typical ML algorithms, such as kinds of neural networks, decision tree, support vectors machine and Bayesian inference. The applications of various machine learning methods to the properties of nuclear structure (e.g. nuclear masses, charge radii, spins, parities, magnetic moments, $\alpha$- and $\beta$-decay properties, and the properties of excited states), nuclear fission yields, reaction data, reaction mechanism, properties of dense matter and experimental techniques, are also reviewed. 

The great success of ML in low and intermediate energy nuclear physics have been achieved mainly in following aspects: 1) the data predictions and evaluations, which is much better than the accuracies of existing models even in the unknown region near or not far from the known region; 2) the discovery and inference of physics from the data which is much better than the traditional statistical tools in high-dimensional parameter space; and 3) the design of experimental detectors, particle identification, and experiments simulations. 

Furthermore, with the progress in ML algorithms, computing power and the corresponding availability of large datasets in nuclear physics, one can expect that the ML will play more important roles in both nuclear theories and experiments. In theory, one can use the machine learning to build nuclear model, such as learning the energy density functional with the machine learning or solving many-body equation for nuclear structure and reactions for enhancing the predictive of theory model. ML can help us to exploit the maximum values of imperfect nuclear fission data. In experiment, particle accelerator facilities and nuclear physics instrumentation face a variety of technical challenges in simulations, control, data acquisition, and analysis that artificial intelligence holds promise to address. Thus, the projects aim to optimize the overall performance of complex accelerator and detector systems for nuclear physics using advanced ML methods are needed.

In the near future, the following directions should be at least  focused for improving the performance of machine learning and discovering the new physics in theories and experiments. The first direction is to involve the physics in the machine learning methods, that was also named as physics informed or physics guided in other literature, to  develop new ML algorithms and overcome the nonphysical predictive and big-data requirements. The second direction is to use the machine learning on quantum computer, which can also boost machine learning in treating extremely large database~\cite{Havlicek2019,LiWeikang2022}.

\section*{Acknowledgements}
We thank the supports of the National Natural Science Foundation of China with Nos. 11875070, 11875323, 12275359, 11875125, 12147219, U2032145, 11705163, 11790320, 11790323, 11790325, 11975032, 11835001, 11935001, 11890710,  12147101, and 11961141003, the National Key R\&D Program of China under Grant No. 2018 YFA0404404, No. 2018YFA0404403, No. 2020YFE0202001, the Continuous Basic Scientific Research Project (No. WDJC-2019-13), the Continuous Basic Scientific Research Project (No. WDJC-2019-13), the funding of China Institute of Atomic Energy (No. YZ222407001301), the Leading Innovation Project of the CNNC under Grant Nos. LC192209000701 and  LC202309000201,  and by Guangdong Major Project of Basic and Applied Basic Research No. 2020B0301030008.

\end{document}